\documentclass[twocolumn,showpacs,superscriptaddress]{revtex4}
\usepackage[latin1]{inputenc}
\usepackage[T1]{fontenc}
\usepackage{graphicx}
\usepackage[dvips]{epsfig}
\usepackage{amsmath}
\usepackage{amssymb}
\usepackage{color}
\usepackage{dcolumn}
\usepackage{slashbox}
\usepackage{braket}
\usepackage{bbm}
\usepackage{mathrsfs}
\usepackage{subfigure}

\newcommand{\Ec}{E_{\text{c}}}

\definecolor{gold}{rgb}{1,0.75,0}

\newcommand{\colH}[1]{{#1}}
\newcommand{\colG}[1]{{#1}}

\begin{document}

\title{\colG{Schwinger pair production in space- and time-dependent electric fields: Relating the Wigner formalism to quantum kinetic theory}}
\author{F.~Hebenstreit}
\author{R.~Alkofer}
\affiliation{Institut f\"ur Physik, Karl-Franzens Universit\"at Graz, 
A-8010 Graz, Austria}
\author{H.~Gies}
\affiliation{Theoretisch-Physikalisches Institut, Friedrich-Schiller
  Universit\"at Jena, \\
  \& Helmholtz-Institut Jena, D-07743 Jena, Germany} 
\date{\today}

\begin{abstract}
  The non-perturbative electron-positron pair production (Schwinger effect) is
  considered for space- and time-dependent electric fields
  $\vec{E}(\vec{x},t)$. Based on the Dirac-Heisenberg-Wigner (DHW) formalism
  we derive a system of partial differential equations of infinite order for
  the sixteen irreducible components of the Wigner function. In the limit of
  spatially homogeneous fields the Vlasov equation of quantum kinetic theory
  (QKT) is rediscovered.  It is shown that the quantum kinetic formalism can
  be exactly solved in the case of a constant electric field $E(t)=E_0$ and
  the Sauter-type electric field
  $E(t)=E_0\operatorname{sech}^2(t/\tau)$. These analytic solutions translate
  into corresponding expressions within the DHW formalism and allow to discuss
  the effect of higher derivatives. We observe that spatial field variations
  typically exert a strong influence on the \colG{components of the Wigner function for large momenta}
  or \colG{for} late times.
\end{abstract}
\pacs{\colG{
12.20.Ds, 
11.15.Tk} 
}

\maketitle

\section{Introduction}
Pair production in strong external electric fields is in many respects a paradigmatic phenomenon in quantum field theory
\cite{Sauter:1931,Heisenberg:1935,Schwinger:1951nm}. It is nonperturbative in the coupling times the external field strength. It exemplifies the nontrivial properties of the quantum vacuum, as it manifests the instability of the vacuum against the formation of many-body states. In general, it  depends strongly on the spacetime structure of the external field, such that the pair production process is expected to exhibit features of nonlocality\colG{,} final state correlations and real-time dynamics. Moreover, it is a non-equilibrium process in quantum field theory \colG{and as such} belonging perhaps to the least-well understood branch of modern field theory. Whereas pair proliferation is expected to occur at the critical Schwinger field strength $\Ec= m^2/ e\simeq1.3 \times 10^{18} \text{V}/\text{m}$, recent studies have suggested that pair production might become observable already at lower but dynamically modulated field strengths \cite{Schutzhold:2008pz,Dunne:2009gi,Monin:2009aj, Heinzl:2010vg,DiPiazza:2009py,Bulanov:2010ei,Baier:2009it}. These estimates of the required field strengths indicate that pair production might already become accessible at future high-intensity laser systems such as the extreme light infrastructure ELI \cite{ELI,Tajima:2009,Dunne:2008kc} \colG{or the European XFEL \cite{XFEL,Ringwald:2001ib}}.

Computing pair production in a complicated \colG{space- and time-dependent}
field such as a high-intensity pulse is by no means straightforward. Many
different theoretical methods, such as the propertime method
\cite{Schwinger:1951nm}, WKB techniques
\cite{Brezin:1970xf,Popov:1972,Popov:1973az,Piazza:2004sv,Dumlu:2010ua}, the
Schr\"odinger-Functional approach \cite{Hallin:1994ad}, functional techniques
\cite{Fried:2001ur,Avan:2002dn}, quantum kinetic equations
\cite{Smolyansky:1997fc,Schmidt:1998vi,
  Smolyansky:1997ji,Kluger:1998bm,Alkofer:2001ik,Hebenstreit:2008ae}, being
also closely related to scattering techniques \cite{Dumlu:2009rr}, various
instanton techniques
\cite{Affleck:1981bma,Kim:2000un,Dunne:2005sx,Dunne:2006st,Dietrich:2007vw},
Borel summation \cite{Dunne:1999uy}, propagator constructions
\cite{Dietrich:2003qf}, and worldline numerics \cite{Gies:2005bz} have been
developed to study pair production in external fields. Most of those
approaches 
have \colG{only} been applied to one-dimensional \colG{temporal} or spatial
inhomogeneities, see \cite{Dunne:2006ur} for the only true multidimensional
case.  Also, finite-temperature contributions have been determined which under
the assumption of local thermal equilibrium first occur at the two-loop level
\cite{Gies:1999vb,Dittrich:2000zu}. \colH{For thermal pair production from
  more general initial states, see
  \cite{Gavrilov:2007hq,Gavrilov:2006jb,Monin:2008td}.}

For both, a profound understanding of the phenomenon as well as reliable
quantitative predictions for realistic cases, a formalism that can deal with
arbitrary space- and time-dependent fields is urgently required. This is also
stressed by recent observations of characteristic and potentially easy to
detect signatures of pair production in the momentum distribution of the pairs
which has turned out surprisingly sensitive to the sub-cycle structure of
high-intensity pulses \cite{Hebenstreit:2009km,Dumlu:2010vv}, also exhibiting
information about the quantum statistics of the particles involved
\cite{Hebenstreit:2009uy,PauchyHwang:2009rz}. Such a formalism based on
suitable real-time correlation functions is indeed available and has already
been studied in the context of pair production
\cite{BialynickiBirula:1991tx}. The present work is devoted to exploring this
DHW formalism, putting it into the context also of other work such as quantum
kinetic equations, and performing first systematic studies with the aid of
both exactly soluble cases and within approximative schemes.

This paper is organized as follows: In Sec.~\ref{sec2} we briefly review the
Dirac-Heisenberg-Wigner (DHW) formalism, adopting already a notation which
will prove to be advantageous in the following. We describe how the Quantum
Kinetic Theory (QKT) emerges as a specific limit of the DHW formalism and
present some analytical solutions. In Sec.~\ref{sec3} we introduce a
derivative expansion and discuss its region of validity. In Sec.~\ref{sec4} we
conclude and provide an outlook. Details about the QKT are summarized in
App.~\ref{appa}. The analytical results for the irreducible components of the
Wigner function in the constant electric field and Sauter-type electric field
are given in App.~\ref{appb}.

\section{The equal-time DHW formalism}
\label{sec2}

A classical statistical one-particle system is described by probability
distributions $\mathcal{F}(\vec{x},\vec{p};t)$ in $6$-dimensional phase space
$\{\vec{x},\vec{p}\}$. The generalization for a relativistic quantum field
theory is obtained by choosing an appropriate density operator and performing
a Wigner transformation to $8$-dimensional phase space $\{x^\mu,p^\mu\}$. The
corresponding Wigner operator $\hat{\mathcal{W}}(x,p)$ is manifestly Lorentz
covariant but the associated Wigner function
$\langle\Omega|\hat{\mathcal{W}}(x,p)|\Omega\rangle$ may not have a clear
physical interpretation \cite{Vasak:1987um, Elze:1989un}.

Alternatively, one may drop the manifest Lorentz covariance in favor of a
canonical time evolution from the beginning and start with an equal-time
density operator. The corresponding Wigner operator
$\hat{\mathcal{W}}(\vec{x},\vec{p};t)$ is then defined in $6$-dimensional
phase space $\{\vec{x},\vec{p}\}$. It is an advantage of this approach that
the Wigner function $\langle\Omega|\hat{\mathcal{W}}
(\vec{x},\vec{p};t)|\Omega\rangle$ might be interpreted as quasi-probability
distribution in analogy to classical physics. It is an additional benefit
that the equation of motion might be formulated as initial value problem
\cite{BialynickiBirula:1991tx, Best:1993wq}. Alternatively, one could also
start with the Lorentz covariant formulation and switch to the equal-time
formulation by performing an energy integral over $p_0$
\cite{Zhuang:1995pd,Zhuang:1998kv,Ochs:1998qj}.

We will adopt the equal-time formulation throughout this paper. Due to the
fact that we have dropped manifest Lorentz covariance anyway, we will also fix
the gauge from the beginning. We will choose the temporal gauge $A_0=0$
throughout, such that the electric and magnetic fields are calculable from the
vector potential $\vec{A}(\vec{x},t)$ according to:
\begin{equation}
  \vec{E}(\vec{x},t)=-\partial_t\vec{A}(\vec{x},t) \quad , \quad \vec{B}(\vec{x},t)=\vec{\nabla}\times\vec{A}(\vec{x},t) 
  \  .
\end{equation}

\subsection{Derivation of the DHW formalism}
\label{sec2a}

In this section we define the equal-time Wigner operator
$\hat{\mathcal{W}}(\vec{x},\vec{p};t)$ in the presence of an external
electromagnetic field. By applying a Hartree approximation for the
electromagnetic field, i.e. treating it as C-number field instead of a \colG{operator-valued quantum field}, we are able to derive the equation of motion
for the corresponding Wigner function
$\langle\Omega|\hat{\mathcal{W}}(\vec{x},\vec{p};t)|\Omega\rangle$.

For this, we consider the following equal-time density operator of two Dirac
field operators in the Heisenberg picture:
\begin{eqnarray}
  \label{sec2a:dens_op}
  &\hat{\mathcal{C}}(\vec{x}_1,\vec{x}_2;t)\equiv
  e^{-ie\int_{\vec{x_2}}^{\vec{x_1}}{\vec{A}(\vec{x'},t)\cdot\vec{dx'}}}\left[\Psi(\vec{x_1},t),\bar{\Psi}(\vec{x_2},t)\right] \, , \quad
\end{eqnarray} 
where we have dropped the Lorentz indices for simplicity. Here we choose the
equal-time commutator, since the equal-time anticommutator is trivially
fulfilled for spinor fields. Additionally, in order to preserve gauge
invariance we include a Wilson line factor with an integral of the vector
potential over a straight line. In fact, the choice of the integration path is
not unique, but the present choice \colG{will allow for introducing a properly defined}
kinetic momentum variable $\vec{p}$. In terms of the center-of-mass coordinates
$\vec{x}=\frac{1}{2}(\vec{x_1}+\vec{x_2})$ and $\vec{s}=\vec{x}_1-\vec{x_2}$, it reads:
\begin{equation}
  \hat{\mathcal{C}}(\vec{x};\vec{s};t)=e^{-ie\int_{-1/2}^{1/2}{\vec{A}(\vec{x}+\lambda\vec{s},t)\cdot\vec{s}d\lambda}} [\Psi(\vec{x}+\tfrac{\vec{s}}{2},t),\bar{\Psi}(\vec{x}-\tfrac{\vec{s}}{2},t)]{.}  
\end{equation}
The Wigner operator is then defined as the Fourier transform of
$\hat{\mathcal{C}}(\vec{x};\vec{s};t)$ with respect to the relative coordinate
$\vec{s}$, such that the arguments are the center-of-mass coordinate
$\vec{x}$, the kinetic momentum variable $\vec{p}$ and time $t$:
\begin{eqnarray}
  \label{sec2a:wigner_op}
  &\hat{\mathcal{W}}(\vec{x},\vec{p};t)\equiv-\frac{1}{2}\int{d^3s\,e^{-i\vec{p}\cdot\vec{s}}\, \hat{\mathcal{C}}(\vec{x};\vec{s};t)}\ .&
\end{eqnarray}
Note that if we had defined Eq.~(\ref{sec2a:dens_op}) with
$\Psi^\dagger(\vec{x_2},t)$ instead of $\bar{\Psi}(\vec{x_2},t)$, the
corresponding Wigner operator would have been Hermitian. With our definition,
$\mathcal{W}(\vec{r},\vec{p},t)$ is not Hermitian but transforms like a Dirac
matrix:
\begin{equation}
  \hat{\mathcal{W}}^\dagger(\vec{x},\vec{p};t)=\gamma^0\hat{\mathcal{W}}(\vec{x},\vec{p};t)\gamma^0
  \ .
\end{equation}
In general, the Wigner function is then defined as the 
{expectation value} of the Wigner operator $\langle\Omega|\hat{\mathcal{W}}
(\vec{x},\vec{p};t)|\Omega\rangle$ { with respect to the full interacting
  vacuum}. However, due to the fact that we are mainly interested in
describing Schwinger pair production in the following, we restrict ourselves
to the vacuum state in the Heisenberg picture $|\Omega\rangle=|0\rangle$:
\begin{eqnarray}
  \label{sec2a:wigner_func}
  &\mathcal{W}(\vec{x},\vec{p};t)=-\frac{1}{2}\int{d^3s\,e^{-i\vec{p}\cdot\vec{s}}\,\langle0|\hat{\mathcal{C}}(\vec{x};\vec{s};t)}|0\rangle
  &.
\end{eqnarray}
In order to derive the equation of motion for the Wigner function, we take the
time derivative of Eq.~(\ref{sec2a:wigner_func}) and take the properly gauge
fixed Dirac equation
\begin{equation}
  \left(i\gamma^0\partial_t+i\vec{\gamma}\cdot\left[\vec{\nabla}_{\vec{x}}-ie\vec{A}(\vec{x},t)\right]-m\right)\Psi(\vec{x},t)=0 
\end{equation}
into account. In the course of the derivation we adopt a Hartree approximation
of the electromagnetic field, which should be a good approximation for high
field strengths. This means that we replace the operator-valued
electromagnetic \colG{quantum} field by a C-number electromagnetic field:
\begin{eqnarray}
  \langle0|\colG{\hat{F}}^{\mu\nu}(\vec{x},t)\hat{\mathcal{C}}(\vec{x};\vec{s};t)|0\rangle \longrightarrow F^{\mu\nu}(\vec{x},t)\langle0|\hat{\mathcal{C}}(\vec{x};\vec{s};t)|0\rangle \ .
\end{eqnarray}
Diagrammatically, this approximation corresponds to ignoring higher-loop
radiative corrections. Physically, this implies that final state interactions
as well as mass shift effects are ignored.  This derivation finally yields the
equation of motion for the Wigner function:
\begin{equation}
  \label{sec2a:wigner_func_eom}
  D_t\mathcal{W}=-\frac{1}{2}\vec{D}_{\vec{x}}\left[\gamma^0\vec{\gamma},\mathcal{W}\right]-im\left[\gamma^0,\mathcal{W}\right]-i\vec{P}\left\{\gamma^0\vec{\gamma},\mathcal{W}\right\} 
  , 
\end{equation}
with $D_t$, $\vec{D}_{\vec{x}}$ and $\vec{P}$ denoting the following non-local
pseudo-differential operators:
\begin{alignat}{6}
  &D_t&\ =\ &\ \ \partial_t \ &\ + \ & e  \int_{-1/2}^{1/2}{d\lambda\,\vec{E}(\vec{x}+i\lambda\vec{\nabla}_{\vec{p}},t)\cdot\vec{\nabla}_{\vec{p}}} \ , \nonumber 
  \\
  \label{sec2a:diffop_nonlocal}
  &\vec{D}_{\vec{x}}&\ =\ &\ \vec{\nabla}_{\vec{x}} \ & + \ & e  \int_{-1/2}^{1/2}{d\lambda\,\vec{B}(\vec{x}+i\lambda\vec{\nabla}_{\vec{p}},t)\times\vec{\nabla}_{\vec{p}}} \ , 
  \\
  &\vec{P}&\ =\ &\ \ \vec{p}\ &\ - \ & ie  \int_{-1/2}^{1/2}{d\lambda\,\lambda\,\vec{B}(\vec{x}+i\lambda\vec{\nabla}_{\vec{p}},t)\times\vec{\nabla}_{\vec{p}}} \ . \nonumber
\end{alignat}

As the Wigner function $\mathcal{W}(\vec{x},\vec{p};t)$ is in fact a Dirac
matrix, we may expand it in terms of irreducible components by choosing an
appropriate complete basis set of $4\times4$ matrices
$\{\mathbbm{1},\gamma_5,\gamma^\mu,\gamma^\mu\gamma_5,\sigma^{\mu\nu}\}$. Actually
we choose $16$ real functions (from now on called DHW functions) which
transform under orthochronous Lorentz transformations as scalar
$\mathbbm{s}(\vec{x},\vec{p};t)$, pseudoscalar
$\mathbbm{p}(\vec{x},\vec{p};t)$, vector $\mathbbm{v}_\mu(\vec{x},\vec{p};t)$,
axialvector $\mathbbm{a}_\mu(\vec{x},\vec{p};t)$ and tensor
$\mathbbm{t}_{\mu\nu}(\vec{x},\vec{p};t)$, respectively:
\begin{equation}
  \label{sec2a:wigner_func_exp}
  \mathcal{W}(\vec{x},\vec{p};t)=\frac{1}{4}\left[\mathbbm{1}\mathbbm{s}+i\gamma_5\mathbbm{p}+\gamma^\mu\mathbbm{v}_\mu+
  \gamma^\mu\gamma_5\mathbbm{a}_\mu+\sigma^{\mu\nu}\mathbbm{t}_{\mu\nu}\right] \, .
  \
\end{equation}
Inserting this decomposition into the equation of motion,
Eq.~(\ref{sec2a:wigner_func_eom}), and comparing the coefficients of the basis
matrices, we find a \colG{PDE} system for the 16 DHW
functions. Introducing the compact notation for the tensorial components:
\begin{equation}
  \left(\vec{\mathbbm{t}}_1\right)_i=\mathbbm{t}_{0i}-\mathbbm{t}_{i0} \quad , \quad \left(\vec{\mathbbm{t}}_2\right)_i=\epsilon_{ijk}\mathbbm{t}_{jk}
  \ ,
\end{equation}
this system reads:
\begin{alignat}{8}
  \label{sec2a:dhw1}
  &D_t\,\mathbbm{s}\ &\  &\  &\ -&\ 2\vec{P}\cdot\vec{\mathbbm{t}}_{1}\ &\ =\ &\ &\ & 0\\ 
  \label{sec2a:dhw2}
  &D_t\,\mathbbm{p}\ &\  &\  &\ +&\ 2\vec{P}\cdot\vec{\mathbbm{t}}_{2}\ &\ =\ &\ &2m\ & \mathbbm{a}_0\\ 
  \label{sec2a:dhw3}
  &D_t\,\mathbbm{v}_0\ &\ +&\ \vec{D}_{\vec{x}}\cdot\vec{\mathbbm{v}}\ &\  &\  &\ =\ &\ &\ & 0\\
  \label{sec2a:dhw4}
  &D_t\,\mathbbm{a}_0\ &\ +&\ \vec{D}_{\vec{x}}\cdot\vec{\mathbbm{a}}\ &\  &\  &\ =\ &\ &2m\ & \mathbbm{p}\\
  \label{sec2a:dhw5}
  &D_t\,\vec{\mathbbm{v}}\ &\ + &\ \vec{D}_{\vec{x}}\,\mathbbm{v}_0\ &\ +&\ 2\vec{P}\times\vec{\mathbbm{a}}\ &\ =\ &\ -&2m\ & \vec{\mathbbm{t}}_1 \\
  \label{sec2a:dhw6}
  &D_t\,\vec{\mathbbm{a}}\ &\ +&\ \vec{D}_{\vec{x}}\,\mathbbm{a}_0\ &\ +&\ 2\vec{P}\times\vec{\mathbbm{v}}\ &\ =\ &\ &\ & 0\\
  \label{sec2a:dhw7}
  &D_t\,\vec{\mathbbm{t}}_{1}\ &\ +&\ \vec{D}_{\vec{x}}\times\vec{\mathbbm{t}}_2\ & + &\ 2\vec{P}\,\mathbbm{s}&\ =\ &\ &2m\ & \vec{\mathbbm{v}}\\
  \label{sec2a:dhw8} 
  &D_t\,\vec{\mathbbm{t}}_{2}\ &\ -&\ \vec{D}_{\vec{x}}\times\vec{\mathbbm{t}}_1\ & - &\ 2\vec{P}\,\mathbbm{p}&\ =\ &\ & &0 
\end{alignat}
Note that for spatially homogeneous electromagnetic fields
$F^{\mu\nu}(\vec{x},t)= F^{\mu\nu}(t)$, an enormous simplification occurs as
the non-local operators Eq.~(\ref{sec2a:diffop_nonlocal}) reduce to local
ones:
\begin{alignat}{6}
  &D_t&\ =\ &\ \ \partial_t \ &\ + \ & e \vec{E}(t)\cdot\vec{\nabla}_{\vec{p}}& \ , \nonumber \\ 
  \label{sec2a:diffop_local}
  &\vec{D}_{\vec{x}}&\ =\ &\ \vec{\nabla}_{\vec{x}} \ & + \ & e\vec{B}(t)\times\vec{\nabla}_{\vec{p}} & \ , \\
  &\vec{P}&\ =\ &\ \ \vec{p} & \ . \nonumber
\end{alignat} 

It has been shown previously \cite{BialynickiBirula:1991tx}, that some of the
DHW functions can be given an intuitive interpretation, whereas others do not
have a classical analogue. First, the symmetrized electromagnetic current \colG{$j^\mu(\vec{x},t)=\frac{e}{2}\bra{0}\left[\bar{\Psi}(\vec{x},t),\gamma^\mu\Psi(\vec{x},t)\right]\ket{0}$} is
expressed as:
\begin{equation}
  \colG{j^\mu(\vec{x},t)=e\int{\frac{d^3p}{(2\pi)^3}\mathbbm{v}^\mu(\vec{x},\vec{p};t)}}{.} 
\end{equation}
Additionally, several conservation laws concerning physically observable quantities like the total charge $\mathcal{Q}$, the total energy $\mathcal{E}$, the total linear momentum $\vec{\mathcal{P}}$ and the total angular momentum $\vec{\mathcal{M}}$ are valid,
\begin{equation}
  \frac{d}{dt}\left\{\mathcal{Q};\mathcal{E};\mathcal{\vec{P}};\mathcal{\vec{M}}\right\}=0 \ , 
\end{equation}
with:
\begin{alignat}{6}
  &\mathcal{Q}&\ =\ & e\int{\colG{d\Gamma} \, \mathbbm{v}_0(\vec{x},\vec{p};t)}\,, \\
  &\mathcal{E}&\ =\ & \int{\colG{d\Gamma} \,\left[\vec{p}\cdot\vec{\mathbbm{v}}(\vec{x},\vec{p};t)+ m\,\mathbbm{s}(\vec{x},\vec{p};t)\right]} \nonumber \\
    & & &  \qquad +\frac{1}{2}\int{d^3x\left[|\vec{E}(\vec{x},t)|^2+|\vec{B}(\vec{x},t)|^2\right]},  \\
  &\mathcal{\vec{P}}&\ = & \int{\colG{d\Gamma} \, \vec{p}\,\mathbbm{v}_0(\vec{x},\vec{p};t)} \nonumber \\
    & & & \qquad \qquad +\int{d^3x\vec{E}(\vec{x},t)\times\vec{B}(\vec{x},t)}, \\
  &\vec{\mathcal{M}}&\ =\ & \int{\colG{d\Gamma} \, \left[\vec{x}\times\vec{p}\,\mathbbm{v}_0(\vec{x},\vec{p};t)
  -\frac{1}{2}\vec{\mathbbm{a}}(\vec{x},\vec{p};t)\right]} \nonumber \\
  & & & \qquad \qquad +\int{d^3x\,\vec{x}\times\vec{E}(\vec{x},t)\times\vec{B}(\vec{x},t)} \ ,
\end{alignat}
\colG{with $d\Gamma=d^3x\,d^3p/(2\pi)^3$ denoting the phase-space volume element.} According to these expressions, we may associate $
\mathbbm{s}(\vec{x},\vec{p};t)$ with a {mass density, $
  \mathbbm{v}_0(\vec{x},\vec{p};t)$ with a charge density,
  $\vec{\mathbbm{v}}(\vec{x},\vec{p};t)$ with a current density and
  $\vec{\mathbbm{a}}(\vec{x},\vec{p};t)$ with a spin density{.} Another important
  conservation law concerns the norm of the Wigner function itself:
\begin{equation}
  \frac{d}{dt}\int{\colG{d\Gamma}\,\operatorname{Tr}\left[\mathcal{W}(\vec{x},\vec{p};t)\mathcal{W}^\dagger(\vec{x},\vec{p};t)\right]}=0 \ ,
\end{equation}
which translates into a conservation law for the 16 DHW functions.

\subsection{Quantum kinetic theory (QKT) as limit of the DHW formalism}
\label{sec2b}

In this subsection we show that the DHW formalism in the case of a spatially
homogeneous, time-dependent electric field $\vec{E}(\vec{x},t)=E(t)\vec{e}_3$
and vanishing magnetic field $\vec{B}(\vec{x},t)=0$ yields the well-known
Vlasov equation of QKT for Schwinger pair production
\cite{Smolyansky:1997fc,Schmidt:1998vi, Kluger:1998bm}. For this, we first
calculate the Wigner function for pure vacuum to obtain appropriate initial
conditions. In a second step we simplify the PDE system (\ref{sec2a:dhw1}) --
(\ref{sec2a:dhw8}) to an ODE system \cite{BialynickiBirula:1991tx}, which
turns out to be equivalent to the Vlasov equation
\cite{Bloch:1999eu}. \colH{For an analysis of the relation between the Wigner
  function and QKT for several examples of pair production in non-abelian
  fields, see \cite{Levai:2009mn}.}

In order to calculate the Wigner function for pure vacuum $\mathcal{W}_{\mathrm{vac}}(\vec{x},\vec{p};t)$, we consider first the general expression Eq.~(\ref{sec2a:wigner_func}) for vanishing vector potential: $A(\vec{x},t)=0$. We first decompose the Dirac field operator in its Fourier basis
\begin{equation}
  \label{sec2b:fourier_decomp}
  \Psi(\vec{x},t)=\int{\frac{d^3q}{(2\pi)^3}\widetilde{\psi}(\vec{q},t)e^{i\vec{q}\cdot\vec{x}}} \ ,
\end{equation}
and introduce a decomposition in terms of anti-commuting creation/annihilation
operators as well as four-spinors
\begin{equation}
  \label{sec2b:quantization}
  \widetilde{\psi}(\vec{q},t)=\sum_{s}{\widetilde{u_s}(\vec{q},t)a_{s}(\vec{q})+\widetilde{v_s}(-\vec{q},t)b^\dagger_{s}(-\vec{q})} \ .
\end{equation}
Evaluating the vacuum expectation value and taking advantage of the four-spinor completeness relations, we finally obtain for the vacuum Wigner function:
\begin{equation}
  \mathcal{W}_{\mathrm{vac}}(\vec{x},\vec{p};t)=-\tfrac{1}{2\omega(\vec{p})}\left[\mathbbm{1}m-\vec{\gamma}\cdot\vec{p}\right] \ ,
\end{equation}
with $\omega(\vec{p})=\sqrt{m^2+\vec{p}^2}$. Comparing this expression with
Eq.~(\ref{sec2a:wigner_func_exp}), we immediately see that (a) in the pure
vacuum only $4$ DHW functions do not vanish and (b) these vacuum
functions do not depend on $\vec{x}$ and $t$:
\begin{eqnarray}
  \label{sec2b:vacuum_s}
  \mathbbm{s}_{\mathrm{vac}}(\vec{p})&=&-\tfrac{2m}{\omega(\vec{p})}\ , \\ 
  \label{sec2b:vacuum_v}
  \mathbbm{\vec{v}}_{\mathrm{vac}}(\vec{p})&=&-\tfrac{2\vec{p}}{\omega(\vec{p})}\ .
\end{eqnarray}

After fixing the vacuum initial conditions, we consider next the PDE system
Eq.~(\ref{sec2a:dhw1}) -- (\ref{sec2a:dhw8}) for $\vec{E}(\vec{x},t)=
E(t)\vec{e}_3$ and $\vec{B}(\vec{x},t)=0$ in more detail: Due to spatial
homogeneity, the DHW functions do not depend on the variable $\vec{x}$ and
hence all spatial derivatives vanish. As an immediate consequence,
$\mathbbm{v}_0(\vec{p};t)$ decouples completely. Additionally, due to the fact
that the DHW functions
$\{\mathbbm{p},\mathbbm{a}_0,\vec{\mathbbm{t}}_2\}(\vec{p};t)$ are subject to
a closed set of equations which does not couple to the \colG{non-vanishing} vacuum initial
conditions, these functions have to vanish as well. As a consequence, the PDE
system for former $16$ DHW functions reduces to a PDE system for the remaining
$10$ DHW functions $\vec{\mathbbm{w}}(\vec{p};t)\equiv(\mathbbm{s},
\vec{\mathbbm{v}},\vec{\mathbbm{a}},\vec{\mathbbm{t}}_1)(\vec{p};t)$:
\begin{equation}
  \label{sec2b:dhw_pde}
 \left[\partial_t+eE(t)\partial_{p_3}\right]\vec{\mathbbm{w}}(\vec{p};t)=\mathcal{M}(\vec{p})\vec{\mathbbm{w}}(\vec{p};t)
  \ .
\end{equation}
Here, $\vec{\mathbbm{w}}(\vec{p};t)$ is a column vector and
$\mathcal{M}(\vec{p})$ is the following $10\times10$ matrix:
\begin{equation}
 \mathcal{M}(\vec{p})=\left(\begin{array}{cccc}0&0&0&2\vec{p}\,^{\mathrm{T}}\\0&0&-2\vec{p}^\times&-2m\\0&-2\vec{p}^\times&0&0\\-2\vec{p}&2m&0&0\end{array}\right) \ ,
\end{equation}
with
\begin{equation}
  \vec{p}^\times=\left(\begin{array}{ccc}0&-p_3&p_2\\p_3&0&-p_1\\-p_2&p_1&0\end{array}\right) \ .
\end{equation}

The PDE system Eq.~(\ref{sec2b:dhw_pde}) will be simplified by applying the
method of characteristics. We introduce a new parameter $\alpha$ and assume
that the originally independent variables depend on this new parameter:
\begin{equation}
  \vec{p}=\vec{\pi}(\alpha) \quad \mathrm{and} \quad t=\tau(\alpha) \ .
\end{equation}
Imposing the following equality for any function $\mathcal{F}(\vec{p};t)$
depending on the former independent variables $\vec{p}$ and $t$:
\begin{equation}
  \left[\tfrac{\partial}{\partial t}+eE(t)\tfrac{\partial}{\partial p_3}\right]\mathcal{F}(\vec{p};t) \stackrel{!}{=}\tfrac{d}{d\alpha}\mathcal{F}(\vec{\pi}(\alpha),\tau(\alpha)) \ ,
\end{equation}
we find $\alpha=\tau=t$ and
$\vec{\pi}(\vec{q},t)=\vec{q}-eA(t)\vec{e}_3$. Note that
$\vec{\pi}(\vec{q},t)$ denotes the time-dependent kinetic momentum on a
trajectory, whereas $\vec{q}$, which serves as an integration constant in
the method of characteristics, corresponds to the canonical
momentum. Additionally, we still have the notion of a phase-space kinetic
momentum $\vec{p}$. These three types of momenta have to be clearly
distinguished in the following. To be consistent throughout this paper, we
always denote:
\begin{alignat}{3}
  & \vec{p} & \qquad & \mathrm{kinetic \ momentum \ in \ phase\ space} 
 \ ,\nonumber \\
  & \vec{q} & \qquad & \mathrm{canonical \ momentum} \ ,\nonumber \\
  & \vec{\pi}(\vec{q},t) & \qquad & \mathrm{kinetic \ momentum \ on \ a \ trajectory} \ . \nonumber 
\end{alignat}
On the one hand, any function defined in phase space possesses only an
explicit time dependence and will {henceforth} be denoted  by
$\mathcal{F}(\vec{p};t)$. On the other hand, functions depending on the
time-dependent kinetic momentum $\vec{\pi}(\vec{q},t)$ show both an explicit
and an implicit time dependence and will be denoted by
$\mathcal{\widetilde{F}}(\vec{q},t)$. 

Formally, the method of characteristics is applied to the PDE system
Eq.~(\ref{sec2b:dhw_pde}) by replacing $\vec{p}$ by $\vec{\pi}(\vec{q},t)$, \colG{such that the relation between the phase-space DHW functions and the DHW functions on a trajectory reads:}
\colG{
\begin{eqnarray}
  \vec{\widetilde{\mathbbm{w}}}(\vec{q},t)&=&\vec{\mathbbm{w}}(\vec{p};t)|_{\vec{p}\to\vec{q}-e\vec{A}(t)} \\ 
  \label{sec2b:dhw_qkt}
  \vec{\mathbbm{w}}(\vec{p};t)&=&\vec{\widetilde{\mathbbm{w}}}(\vec{q},t)|_{\vec{q}\to\vec{p}+e\vec{A}(t)}
\end{eqnarray}}
\noindent Consequently, the PDE system Eq.~(\ref{sec2b:dhw_pde}) becomes an ODE system,
with the former time-independent matrix $\mathcal{M}(\vec{p})$ becoming a
time-dependent quantity $\widetilde{\mathcal{M}}(\vec{q},t)$:
\begin{equation}
   \label{sec2b:dhw_ode}
   \frac{d}{dt}\vec{\widetilde{\mathbbm{w}}}(\vec{q},t)=\widetilde{\mathcal{M}}(\vec{q},t) \vec{\widetilde{\mathbbm{w}}}(\vec{q},t) \ .
\end{equation}
In order to proceed, we seek an appropriate basis to span
$\vec{\widetilde{\mathbbm{w}}}(\vec{q},t)$, such that
Eq.~(\ref{sec2b:dhw_ode}) reduces to a simple form:
\begin{equation}
  \label{sec2b:basis_general}
  \vec{\widetilde{\mathbbm{w}}}(\vec{q},t)=-2\sum_{i=1}^{10}{\widetilde{\chi}^{i}(\vec{q},t)\vec{\widetilde{\mathbbm{e}}}_{i}(\vec{q},t)} \ ,
\end{equation}
with the factor $-2$ chosen for later convenience. To this end, we exploit the
vacuum initial conditions Eq.~(\ref{sec2b:vacuum_s}) -- (\ref{sec2b:vacuum_v})
and choose the first basis vector $\vec{\widetilde{\mathbbm{e}}}_{1}
(\vec{q},t)$ such that in pure vacuum the first coefficient
$\widetilde{\chi}^{1}_{\mathrm{vac}}(\vec{q},t_\mathrm{vac})=1$, whereas all
other coefficients
$\widetilde{\chi}^{i}_{\mathrm{vac}}(\vec{q},t_\mathrm{vac})$
vanish. Consequently, we find a subset of basis vectors:
\begin{equation}
  \label{sec2b:basis_vec1}
  \vec{\widetilde{\mathbbm{e}}}_{1}(\vec{q},t)  =   \frac{1}{\widetilde{\omega}(\vec{q},t)}\left(\begin{array}{c}m\\\vec{\pi}(\vec{q},t)\\\vec{0}\\\vec{0}\end{array}\right) \ , \nonumber
\end{equation}
\begin{equation}
  \label{sec2b:basis_vec2}
  \vec{\widetilde{\mathbbm{e}}}_{2}(\vec{q},t)  =   \frac{1}{\epsilon_\perp\widetilde{\omega}(\vec{q},t)}\left(\begin{array}{c}m\,\pi_3(q_3,t)\\\vec{\pi}(\vec{q},t)\,\pi_3(q_3,t)-\widetilde{\omega}^2(\vec{q},t)\vec{e}_3\\\vec{0}\\\vec{0}\end{array}\right) \ ,
\end{equation}
\begin{equation}
  \label{sec2b:basis_vec3}
  \vec{\widetilde{\mathbbm{e}}}_{3}(\vec{q},t)  =   \frac{1}{\epsilon_\perp}\left(\begin{array}{c}0\\\vec{0}\\\vec{\pi}(\vec{q},t)\times\vec{e}_3\\-m\vec{e}_3\end{array}\right) \ , \nonumber
\end{equation}
with $\epsilon_\perp=\sqrt{m^2+\vec{q}_\perp^2}$ and $\widetilde{\omega}(\vec{q},t)=\sqrt{\epsilon_\perp^2+\pi_3^2(q_3,t)}$, which form an orthonormalized, complete set:
\begin{equation}
  \label{sec2b:basis_relation1}
  \widetilde{M}(\vec{q},t)\left\{\begin{array}{c}\vec{\widetilde{\mathbbm{e}}}_{1}\\\vec{\widetilde{\mathbbm{e}}}_{2}\\\vec{\widetilde{\mathbbm{e}}}_{3}\end{array}\right\}(\vec{q},t)=  2\widetilde{\omega}(\vec{q},t)\left\{\begin{array}{c}\vec{0}\\\vec{\widetilde{\mathbbm{e}}}_{3}\\-\vec{\widetilde{\mathbbm{e}}}_{2}\end{array}\right\}(\vec{q},t) \ , 
\end{equation}
\begin{equation}
  \label{sec2b:basis_relation2}
  \frac{d}{dt}\left\{\begin{array}{c}\vec{\widetilde{\mathbbm{e}}}_{1}\\\vec{\widetilde{\mathbbm{e}}}_{2}\\\vec{\widetilde{\mathbbm{e}}}_{3}\end{array}\right\}(\vec{q},t)=
  -\frac{eE(t)\epsilon_\perp}{\widetilde{\omega}^2(\vec{q},t)}\left\{\begin{array}{c}\vec{\widetilde{\mathbbm{e}}}_{2}\\-\vec{\widetilde{\mathbbm{e}}}_{1}\\\vec{0}\end{array}\right\}(\vec{q},t) \ .
\end{equation}
As a consequence, only the coefficients $\widetilde{\chi}^{i=\{1,2,3\}}(\vec{q},t)$ couple to the initial vacuum state whereas all other coefficients $\widetilde{\chi}^{i}(\vec{q},t)$ \colG{vanish}. This means that $\vec{\widetilde{\mathbbm{w}}}(\vec{q},t)$ is fully characterized by:
\begin{equation}
  \label{sec2b:basis_simplified}
  \vec{\widetilde{\mathbbm{w}}}(\vec{q},t)=-2\sum_{i=1}^{3}{\widetilde{\chi}^{i}(\vec{q},t)\vec{\widetilde{\mathbbm{e}}}_{i}(\vec{q},t)} \ .
\end{equation}

Next we introduce $\widetilde{f}(\vec{q},t)=1-\widetilde{\chi}^1(\vec{q},t)$
parametrizing the deviation from the vacuum state, such that in pure vacuum
$\widetilde{f}_\mathrm{vac}(\vec{q},t_\mathrm{vac})=0$. Additionally, we
define:
\begin{equation}
  \label{sec2b:q_function}
  \widetilde{Q}(\vec{q},t)=\frac{eE(t)\epsilon_\perp}{\widetilde{\omega}^2(\vec{q},t)} \ .
\end{equation}
If we consider the ODE system Eq.~(\ref{sec2b:dhw_ode}) together with the relations Eq.~(\ref{sec2b:basis_relation1}) -- (\ref{sec2b:basis_relation2}), we obtain:
\begin{eqnarray}
  \label{sec2b:dhw_qkt1}
  \frac{d}{dt}\widetilde{f}(\vec{q},t)&=&\widetilde{Q}(\vec{q},t)\,\widetilde{\chi}^2(\vec{q},t) \ , 
  \\
  \label{sec2b:dhw_qkt2}
  \frac{d}{dt}\widetilde{\chi}^2(\vec{q},t)&=&\widetilde{Q}(\vec{q},t)[1-\widetilde{f}(\vec{q},t)]-2\widetilde{\omega}(\vec{q},t)\,\widetilde{\chi}^3(\vec{q},t) \ , \qquad
  \\
  \label{sec2b:dhw_qkt3}
  \frac{d}{dt}\widetilde{\chi}^3(\vec{q},t)&=&2\widetilde{\omega}(\vec{q},t)\,\widetilde{\chi}^2(\vec{q},t) \ , 
\end{eqnarray}
together with vacuum initial conditions
$\widetilde{f}_\mathrm{vac}(\vec{q},t_\mathrm{vac})=
\widetilde{\chi}^2_\mathrm{vac}
(\vec{q},t_\mathrm{vac})=\widetilde{\chi}^3_\mathrm{vac}(\vec{q},t_\mathrm{vac})=0$. This
ODE system is nothing but the well-known Vlasov equation of QKT in its
differential form \cite{Bloch:1999eu} (cf. also \colG{App.} \ref{appa}). Note that
$\widetilde{f}(\vec{q},t)$ denotes the single-particle momentum distribution
function in quantum kinetic theory. Thus, the DHW formalism in the presence of
a spatially homogeneous, time-dependent electric field
$\vec{E}(\vec{x},t)=E(t)\vec{e}_3$ is completely equivalent to QKT. However,
the DHW formalism is the more general approach since it allows for any time-
and space-dependent electromagnetic field whereas the Vlasov equation is
restricted to spatially homogeneous, time-dependent electric fields.

For some special cases, an exact solution of QKT can be found (see
Sec.~\ref{sec2c}), such that we are able to calculate the DHW functions as
well. To this end one uses the representation
Eq.~(\ref{sec2b:basis_simplified}) and projects back to phase-space
Eq.~(\ref{sec2b:dhw_qkt}). 
\colG{For Schwinger pair production in
spatially homogeneous, time-dependent electric fields, for instance, one finds that only
$7$ of the possible 16 DHW functions contribute}:
\begin{alignat}{8}
  \label{sec2b:dhw_s}
  &\mathbbm{s}(\vec{p};t)& \ = \ & - & \frac{2m}{\omega(\vec{p})}&\chi^{1}(\vec{p};t)& \ - \ & 
  \frac{2m\,p_3}{\epsilon_\perp\omega(\vec{p})}&\chi^{2}(\vec{p};t)& \ , \\
  \label{sec2b:dhw_vperp} 
  &\vec{\mathbbm{v}}_\perp(\vec{p};t)& \ = \ &  - & \frac{2\vec{p}_\perp}{\omega(\vec{p})}&\chi^{1}(\vec{p};t)& \ - \ & \frac{2\vec{p}_\perp p_3}{\epsilon_\perp\omega(\vec{p})}&\chi^{2}(\vec{p};t)& \ , \\  
  \label{sec2b:dhw_v3} 
  &\mathbbm{v}_3(\vec{p};t)& \ = \ & - & \frac{2p_3}{\omega(\vec{p})}&\chi^{1}(\vec{p};t)& \ + \ & \ \ \ \frac{2\epsilon_\perp}{\omega(\vec{p})}&\chi^{2}(\vec{p};t)& \ , \\
  \label{sec2b:dhw_a1}
  &\mathbbm{a}_1(\vec{p};t)& \ = \ &  - & \frac{2p_2}{\epsilon_\perp}&\chi^{3}(\vec{p};t)& \ , \\
  \label{sec2b:dhw_a2} 
  &\mathbbm{a}_2(\vec{p};t)& \ = \ &  &  \frac{2p_1}{\epsilon_\perp}&\chi^{3}(\vec{p};t)& \ , \\
  \label{sec2b:dhw_t13}
  &\mathbbm{t}_{1,3}(\vec{p};t)& \ = \ & & \frac{2m}{\epsilon_\perp}&\chi^{3}(\vec{p};t)& \ .
\end{alignat}

\subsection{Exactly solvable electric fields}
\label{sec2c}

In this subsection we derive the analytic expressions for the single-particle
momentum distribution function $\widetilde{f} (\vec{q},t)$ of QKT for both the
constant electric field $E(t)=E_0$ and the Sauter-type electric field
$E(t)=E_0 \operatorname{sech}^2(t/\tau)$. The construction of the solution can
be oriented along the lines of QKT, cf. App.~\ref{appa}: We first
seek an analytic solution for $\widetilde{g}^{(+)}(\vec{q},t)$ of
Eq.~(\ref{appa:eom_g}) and determine its normalization such that it coincides
at $t_\mathrm{vac}\to-\infty$ with $\widetilde{G}^{(+)}(\vec{q},t)$ defined in
Eq.~(\ref{appa:g_adiabatic}). According to Eq.~(\ref{appa:bogoliubov_b}), we
are then able to calculate the Bogoliubov coefficient
$\widetilde{\beta}(\vec{q},t)$ and, consequently, the single-particle
momentum distribution function $\widetilde{f} (\vec{q},t)$ according to
Eq.~(\ref{appa:dist_function}).

As soon as this solution is know, we are able to calculate the non-vanishing coefficients $\widetilde{\chi} ^{i=\{2,3\}}(\vec{q},t)$ according to Eq.~(\ref{sec2b:dhw_qkt1}) -- (\ref{sec2b:dhw_qkt3}). As an immediate consequence of Eq.~(\ref{sec2b:dhw_s}) -- (\ref{sec2b:dhw_t13}), all the non-vanishing DHW functions can be calculated as well.

\subsubsection{Constant electric field}

A constant electric field $E(t)=E_0$ might be represented by the vector potential:
\begin{equation}
  A(t)=-E_0t \ ,
\end{equation}
such that Eq.~(\ref{appa:eom_g}) reads:
\begin{equation}
  \label{sec2c:1_eom}
  \left(\partial_t^2+\epsilon_\perp^2+(q_3+eE_0t)^2+ieE_0\right)\widetilde{g}(q_3,t)=0 \ .
\end{equation}
Note that for notational simplicity we do not explicitly indicate the dependence on the orthogonal canonical momentum via $\epsilon_\perp^2=m^2+\vec{q}^2_\perp$. Dynamically, $q_3$ is the only relevant paramater such that the situation becomes effectively \colG{one}-dimensional. Introducing the dimensionless parameter $\eta=\epsilon_\perp^2/eE_0$ and performing the variable transformation:
\begin{eqnarray}
 \label{sec2c:1_vartrans}
 &q_3+eE_0t=\sqrt{\frac{eE_0}{2}}u \ , 
\end{eqnarray}
we see that $\widetilde{g}(u)$ will only depend on one dimensionless variable $u$ due to the linear relation between $q_3$ and $t$. As a consequence, the differential equation Eq.~(\ref{sec2c:1_eom}) turns into the parabolic cylinder differential equation (\cite{Abramowitz}, Chapter 19):
\begin{eqnarray}
  &\left[\partial_u^2+\frac{1}{4}u^2+\frac{1}{2}(i+\eta)\right]\widetilde{g}(u)=0 \ .
\end{eqnarray}
This second order differential equation has two standard solutions which are given by:
\begin{alignat}{4}
  \label{sec2c:1_g_plus}
  &\widetilde{g}^{(+)}(u)& \ = \ & N^{(+)}D_{-1+i\eta/2}(-ue^{-i\frac{\pi}{4}})\ , & \\
  \label{sec2c:1_g_minus}
  &\widetilde{g}^{(-)}(u)& \ = \ & N^{(-)}D_{-i\eta/2}(-ue^{i\frac{\pi}{4}}) \ , & 
\end{alignat}
with $D_{\nu}(z)$ being the parabolic cylinder function and $N^{(\pm)}$ being
normalization factors. For $u\rightarrow-\infty$, these solutions behave
asymptotically as:
\begin{eqnarray}
  &\widetilde{g}^{(+)}(u)  \stackrel{u\to-\infty}{\longrightarrow}   \frac{1}{|u|}e^{i[\frac{u^2}{4}+\frac{\eta}{2}\log(|u|)+\frac{\pi}{4}]}e^{\frac{\pi\eta}{8}} \ , & \\
  &\widetilde{g}^{(-)}(u)  \stackrel{u\to-\infty}{\longrightarrow}   e^{-i[\frac{u^2}{4}+\frac{\eta}{2}\log(|u|)]}e^{\frac{\pi\eta}{8}} \ . \qquad &
\end{eqnarray}
On the other hand, the adiabatic mode functions are given by, cf.
Eq.~(\ref{appa:g_adiabatic}):
\begin{equation}
  \widetilde{G}^{(\pm)}(u)=\frac{e^{\mp i\Theta(u_{0},u)}}{\sqrt{m^2\epsilon\sqrt{2\eta+u^2}(\sqrt{2\eta+u^2}\mp u)}}
  \ ,
\end{equation}
with:
\begin{equation}
 \epsilon=E_0/\Ec \ .
\end{equation}
The dynamical phase $\Theta(u_0,u)$ can be explicitly calculated as soon as we
fix $u_0$. Here, we choose the symmetric point $u_0=0$, such that \colG{the definite integral}{,}
\begin{equation}
  \Theta(0,u)=\frac{1}{2}\int_{0}^{u}{du'\sqrt{2\eta+u'^2}}{,}
\end{equation}
\colG{yields:}
\begin{equation}
\frac{1}{4}\left[u\sqrt{2\eta+u^2}+2\eta\log\left(u+\sqrt{2\eta+u^2}\right)-\eta\log{2\eta}\right] \ . \\ 
\end{equation}
\newline
\colG{Consequently}, we fix the normalization constants $N^{(\pm)}$ such that
$\widetilde{g}^{(\pm)}(u\to-\infty)=\widetilde{G}^{(\pm)}(u\to-\infty)$:
\begin{eqnarray} &N^{(+)} =
\frac{1}{\sqrt{2m^2\epsilon}}e^{i\left[\frac{\eta}{4}[1+\log(2/\eta)]-\frac{\pi}{4}\right]}e^{-\frac{\pi\eta}{8}}\
, & \\ &N^{(-)} =
\frac{1}{\sqrt{m^2\epsilon\,\eta}}e^{-i\frac{\eta}{4}[1+\log(2/\eta)]}e^{-\frac{\pi\eta}{8}}
\ .&
\end{eqnarray}
According to Eq.~(\ref{appa:dist_function}) and Eq.~(\ref{appa:bogoliubov_b}),
we are \colG{then} able to calculate the single-particle momentum distribution
function. Taking into account the general relation for parabolic cylinder
functions:
\begin{eqnarray} &\partial_zD_{\nu}(z)=\frac{1}{2}zD_{\nu}(z)-D_{1+\nu}(z) \
,&
\end{eqnarray}
we finally obtain:
\begin{widetext}
\begin{eqnarray}
  \label{sec2c:1_dist}
&\widetilde{f}(u)=\frac{1}{4}\left(1+\frac{u}{\sqrt{2\eta+u^2}}\right)e^{-\frac{\pi\eta}{4}}
\left|\left(\sqrt{2\eta+u^2}-u\right)D_{-1+i\eta/2}(-ue^{-i\frac{\pi}{4}})-2e^{i\frac{\pi}{4}}D_{i\eta/2}(-ue^{-i\frac{\pi}{4}})\right|^2
\ .
\end{eqnarray}
\end{widetext}
Finally, we may show that we obtain the Schwinger result for the pair
production rate in a constant electric field if we consider the limit
$u\to\infty$. To this end, we take the leading term in the asymptotic
expansion of the parabolic cylinder functions in
Eq.~(\ref{sec2c:1_dist}). Neglecting terms of the order $\mathcal{O}(u^{-1})$,
they are given by:
\begin{eqnarray}
&D_{-1-i\eta/2}(-u^{i\frac{\pi}{4}})\stackrel{u\to\infty}\longrightarrow\frac{\sqrt{2\pi}}{\Gamma(1+i\eta/2)}e^{i[\frac{u^2}{4}+\frac{\eta}{2}\log(u)]}e^{\frac{-\pi\eta}{8}}{,}
\ \quad & \\
&D_{-i\eta/2}(-u^{i\frac{\pi}{4}})\stackrel{u\to\infty}{\longrightarrow}
e^{-i[\frac{u^2}{4}+\frac{\eta}{2}\log(u)]}e^{-\frac{3\pi\eta}{8}}{,}
\qquad 
\quad \ &
\end{eqnarray}
such that the asymptotic behavior of $\widetilde{f}(u)$ is given by:
\begin{equation} \lim_{u\to\infty}\widetilde{f}(u)= 2e^{-\pi\eta} \ .
\end{equation}
As a consequence, the Schwinger pair production rate per volume and time
$\dot{n}[e^+e^-]$, i.e. the first term in the Schwinger expression for the
vacuum decay probability \cite{Cohen:2008wz,Tanji:2008ku}, is found:
\begin{eqnarray}
&\dot{n}[e^+e^-]=\int{\frac{d^3q}{(2\pi)^3}\partial_t\widetilde{f}(\vec{q},t)}=\frac{e^2E_0^2}{4\pi^3}
e^{-\frac{m^2\pi}{eE_0}} \ , &
\end{eqnarray}
which completes our analytical solution for the constant electric field.

\subsubsection{Sauter-type electric field}
\label{sec:sauter}

The Sauter-type electric field $E(t)=E_0\operatorname{sech}^2(t/\tau)$ might
be represented by the vector potential:
\begin{eqnarray}
&A(t)=-E_0\tau\,\operatorname{tanh}\left(\tfrac{t}{\tau}\right) \ ,
\end{eqnarray}
such that Eq.~(\ref{appa:eom_g}) reads:
\colG{
\begin{eqnarray}
  \label{sec2c:2_eom_1}
&\Big[\partial_t^2+\epsilon_\perp^2+\left(q_3+eE_0\tau\operatorname{tanh}\left(\tfrac{t}{\tau}\right)\right)^2+ \qquad \qquad \qquad \nonumber \\ &\qquad \qquad \qquad \qquad +ieE_0\operatorname{sech}^2(\frac{t}{\tau})\Big]\widetilde{g}(q_3,t)=0
\ .
\end{eqnarray}}
\noindent Again, we only indicate the dependence on $q_3$ whereas the dependence on the
orthogonal canonical momentum will not be denoted explicitly. In the
following, we introduce the dimensionless variable:
\begin{eqnarray}
&u=\tfrac{1}{2}\left[1+\operatorname{tanh}\left(\tfrac{t}{\tau}\right)\right]& \
,
\end{eqnarray}
such that $t\to-\infty$ corresponds to $u\to0$ whereas $t\to\infty$
corresponds to $u\to1$. Additionally, we introduce the Keldysh parameter
$\gamma=1/(m\epsilon\tau)$ and dimensionless momentum variables
$\hat{q}_3=q_3/m$ and $\hat{\kappa}^2=\vec{q}_\perp^2/m^2$, such that the
dimensionless kinetic momentum on the trajectory
$\widehat{\pi}_3(\hat{q}_3,u)$ and the dimensionless energy variable
$\widehat{\omega}(\hat{q}_3,u)$ read:
\begin{eqnarray} &\widehat{\pi}_3(\hat{q}_3,u) =
\frac{q_3+eE_0\tau(2u-1)}{m}=\hat{q}_3+\frac{2u-1}{\gamma}\ , & \\
&\widehat{\omega}^2(\hat{q}_3,u) =
1+\hat{\kappa}^2+\widehat{\pi}^2_3(\hat{q}_3,u) \ .&
\end{eqnarray}
Within these new variables, the differential equation
Eq.~(\ref{sec2c:2_eom_1}) reads:
\colG{
\begin{eqnarray}
  \label{sec2c:2_eom_2}
&\Big[4\gamma^2\epsilon^2u(1-u)\partial_u\{u(1-u)\}\partial_u+\widehat{\omega}^2(\hat{q}_3,u)+\qquad \ \ \nonumber \\
&\qquad \qquad \qquad \qquad +4i\epsilon u(1-u)\Big]\widetilde{g}(\hat{q}_3,u)=0 \ .
\end{eqnarray}}
\noindent In order to solve this differential equation, we apply an ansatz for
$\widetilde{g}(\hat{q}_3,u)$:
\begin{equation}
  \label{sec2c:ansatz} \widetilde{g}(\hat{q}_3,u)=
u^{-i\frac{\widehat{\omega}(\hat{q}_3,0)}{2\gamma\epsilon}}(1-u)^{i\frac{\widehat{\omega}(\hat{q}_3,0)}{2\gamma\epsilon}}\widetilde{h}(\hat{q}_3,u) 
,
\end{equation}
Plugging this ansatz into Eq.~(\ref{sec2c:2_eom_2}) yields the hypergeometric
differential equation (\cite{Abramowitz}, Chapter 15) for
$\widetilde{h}(\hat{q}_3,u)$:

\begin{equation}
\left[u(1-u)\partial_u^2+(\tilde{c}-[\tilde{a}+\tilde{b}+1]u)\partial_u-\tilde{a}\,\tilde{b}\right]\widetilde{h}(\hat{q}_3,u)=0
\ ,
\end{equation}
with
\begin{eqnarray} \tilde{a}(\hat{q}_3)\quad =& \ \ \ \ - \
\frac{i}{\gamma\epsilon}\left(\frac{1}{\gamma}+\frac{\widehat{\omega}(\hat{q}_3,0)}{2}-\frac{\widehat{\omega}(\hat{q}_3,1)}{2}\right)
\ ,& \nonumber \\
  \label{sec2c:2_arguments} \tilde{b}(\hat{q}_3)\quad =&\ \, \,
1+\frac{i}{\gamma\epsilon}\left(\frac{1}{\gamma}-\frac{\widehat{\omega}(\hat{q}_3,0)}{2}+\frac{\widehat{\omega}(\hat{q}_3,1)}{2}\right)
\ ,& \\ \tilde{c}(\hat{q}_3)\quad
=&1-\frac{i}{\gamma\epsilon}\,\widehat{\omega}(\hat{q}_3,0) \ .\qquad \qquad
\qquad & \nonumber
\end{eqnarray}
Note that these parameters do not depend on $u$. The two linearly independent
solutions $\widetilde{h}^{(\pm)}(\hat{q}_3,u)$ in the neighborhood of the
singular point $u=0$ are given by:
\colG{\begin{eqnarray}
  \label{sec2c:2_h_plus}
\widetilde{h}^{(+)}(\hat{q}_3,u)&=&N^{(+)}F(\tilde{a},\tilde{b},\tilde{c};u), \\
  \label{sec2c:2_h_minus}
\widetilde{h}^{(-)}(\hat{q}_3,u)&=&N^{(-)}u^{i\frac{\widehat{\omega}(\hat{q}_3,0)}{\gamma\epsilon}}(1-u)^{-i\frac{\widehat{\omega}(\hat{q}_3,1)}{\gamma\epsilon}}\nonumber\\&&\qquad\qquad
{\times}F(1-\tilde{a},1-\tilde{b},2-\tilde{c};u) \ , \quad
\end{eqnarray}}
\newline
with $F(\tilde{a},\tilde{b},\tilde{c};u)$ denoting the Gauss hypergeometric \colG{function}. Taking
Eq.~(\ref{sec2c:ansatz}) into account, the \colG{asymptotic} behavior of
$\widetilde{g}^{(\pm)}(\hat{q}_3,u)$ for $u\to0$ is:
\begin{eqnarray}
\widetilde{g}^{(+)}(\hat{q}_3,u)&\stackrel{u\to0^+}{\longrightarrow}&e^{-i\frac{\widehat{\omega}(\hat{q}_3,0)}{2\gamma\epsilon}\log(u)}
\ , \\
\widetilde{g}^{(-)}(\hat{q}_3,u)&\stackrel{u\to0^+}{\longrightarrow}&e^{i\frac{\widehat{\omega}(\hat{q}_3,0)}{2\gamma\epsilon}\log(u)}
\ .
\end{eqnarray}
On the other hand, we are again able to give an analytic expression for the
adiabatic mode functions:
\begin{equation} \widetilde{G}^{(\pm)}(\hat{q}_3,u)=\frac{e^{\mp
i\Theta(\hat{q}_3,u_{0},u)}}{\sqrt{2m^2\widehat{\omega}(\hat{q}_3,u)[\widehat{\omega}(\hat{q}_3,u)\mp\widehat{\pi}_3(\hat{q}_3,u)]}}
\ ,
\end{equation}
Again, the dynamical phase $\Theta(\hat{q}_3,u_0,u)$, which is given by the
following integral:
\begin{eqnarray}
\Theta(\hat{q}_3,u_0,u)=&\frac{1}{2\gamma\epsilon}\int_{u_0}^{u}{du'\frac{\sqrt{1+\hat{\kappa}^2+\widehat{\pi}_3^2(\hat{q}_3,u')}}{u'(1-u')}}\
,
\end{eqnarray}
can be analytically calculated. For $u_0\ne\{0,1\}$ and $u\to0$, this phase
splits into a relevant divergent part and an irrelevant regular part
$\Psi(\hat{q}_3,u_0,0)$:
\begin{eqnarray}
&\Theta(\hat{q}_3,u_0,0)=\Psi(\hat{q}_3,u_0,0)+\frac{\widehat{\omega}(\hat{q}_3,0)}{2\gamma\epsilon}\log(u)&
\ ,
\end{eqnarray}
such that the normalization constants $N^{(\pm)}$ is fixed according to
$\widetilde{g}^{(\pm)}(\hat{q}_3,0)= \widetilde{G}^{(\pm)}(\hat{q}_3,0)$:
\begin{eqnarray} &N^{(\pm)} = \frac{e^{\mp
i\Psi(\hat{q}_3,u_0,0)}}{\sqrt{2m^2\widehat{\omega}(\hat{q}_3,0)[\widehat{\omega}(\hat{q}_3,0)\mp\widehat{\pi}_3(\hat{q}_3,0)]}}
\ ,
\end{eqnarray}
Again, the single-particle momentum distribution function is calculated
according to Eq.~(\ref{appa:dist_function}) and Eq.~(\ref{appa:bogoliubov_b}):
\begin{widetext}
\begin{eqnarray}
  \label{sec2c:2_dist}
  &\widetilde{f}(\hat{q}_3,u)=\widetilde{N}_f(\hat{q}_3)\left(1+\frac{\widehat{\pi}_3(\hat{q}_3,u)}{\widehat{\omega}_3(\hat{q}_3,u)}\right)\left|\left[\widehat{\omega}(\hat{q}_3,u)-(1-u)\widehat{\omega}(\hat{q}_3,0)-u\widehat{\omega}(\hat{q}_3,1)\right]F(\tilde{a},\tilde{b},\tilde{c};u)-2i\gamma\epsilon u(1-u)\partial_uF(\tilde{a},\tilde{b},\tilde{c};u)\right|^2 \nonumber \\
 &
\end{eqnarray}
\end{widetext}
with the normalization factor $\widetilde{N}_f(\hat{q}_3)$ being given by:
\begin{eqnarray}
  \label{sec2c:2_normalization}
\widetilde{N}_f(\hat{q}_3)=&\frac{1}{2\widehat{\omega}(\hat{q}_3,0)[\widehat{\omega}(\hat{q}_3,0)-\widehat{\pi}_{3}(\hat{q}_3,0)]}&
.
\end{eqnarray}
and
\begin{eqnarray}
  \label{sec2c:2_hypergeometric}
&\partial_uF(\tilde{a},\tilde{b},\tilde{c};u)=\frac{\tilde{a}\,\tilde{b}}{\tilde{c}}F(1+\tilde{a},1+\tilde{b},1+\tilde{c};z)
\ .
\end{eqnarray}
Similarly to the constant electric field, we may give a simple expression for
the \colG{asymptotic single-particle momentum distribution function}. Applying a linear transformation formula for
hypergeometric functions and considering the asymptotic limit $u\to1$, one
\colG{first }obtains:
\colG{
\begin{eqnarray}
&\widetilde{f}(\hat{q}_3,1)=\qquad \qquad \qquad \qquad \qquad \qquad \qquad \qquad \qquad \quad \nonumber\\ &\frac{2\gamma^2\epsilon^2}{\widehat{\omega}(\hat{q}_3,0)\widehat{\omega}(\hat{q}_3,1)}
\frac{\widehat{\omega}(\hat{q}_3,1)+\hat{\pi}_{3}(\hat{q}_3,1)}{\widehat{\omega}(\hat{q}_3,1)-\hat{\pi}_{3}(\hat{q}_3,1)}
\left|\frac{\tilde{a}\,\tilde{b}}{\tilde{c}}\frac{\Gamma(1+\tilde{c})\Gamma(1+\tilde{a}+\tilde{b}-\tilde{c})}{\Gamma(1+\widetilde{a})\Gamma(1+\tilde{b})}\right|^2 {.}\
\end{eqnarray}}
Applying the transformation formulae for gamma functions
\begin{eqnarray} &\Gamma(1+a)=a\Gamma(a) \quad \mathrm{and} \quad
|\Gamma(1+ib)|^2=\frac{\pi b}{\sinh(\pi b)} \ , \qquad&
\end{eqnarray}
we obtain a relatively simple analytic expression for the asymptotic
single-particle momentum distribution function $\widetilde{f}(\hat{q}_3,1)$:
\colG{
\begin{eqnarray}
&\frac{2\sinh\left(\frac{\pi}{2\gamma\epsilon}\left[\frac{2}{\gamma}+\widehat{\omega}(\hat{q}_3,1)-\widehat{\omega}(\hat{q}_3,0)\right]\right)
\sinh\left(\frac{\pi}{2\gamma\epsilon}\left[\frac{2}{\gamma}-\widehat{\omega}(\hat{q}_3,1)+\widehat{\omega}(\hat{q}_3,0)\right]\right)}
{\sinh\left(\frac{\pi}{\gamma\epsilon}\widehat{\omega}(\hat{q}_3,1)\right)\sinh\left(\frac{\pi}{\gamma\epsilon}\widehat{\omega}(\hat{q}_3,0)\right)}
{.}&
\nonumber \\ 
\end{eqnarray}}

\section{Influence of a small spatial inhomogeneity}
\label{sec3}

In this section{,} we discuss the influence of a small spatial inhomogeneity
along the direction of the time-dependent electric field $\vec{E}(\vec{x},t)$
based on the analytic results for both the constant electric field and the
Sauter-type electric field. In order to estimate the effect of higher
derivatives, we adopt a derivative expansion and determine the ratio between
the first derivative and higher derivatives. Note that we again ignore the
effect of magnetic fields for simplicity, $\vec{B}(\vec{x},t)=0$.

\subsection{Derivative expansion}

We consider a space- and time-dependent electric field:
\begin{equation} \vec{E}(\vec{x},t)=E(t)\left[1+\Delta(x_3)\right]\vec{e}_3 \
,
\end{equation}
where $|\Delta(x_3)|\ll1$ describes a small deviation from the spatially
homogeneous electric field. The equation of motion for the 16 DHW functions
$\mathbbm{\vec{w}}(\vec{x},\vec{p};t)$ {read}:
\begin{equation}
  \label{sec3a:dhw_pde_full}
D_t\mathbbm{\vec{w}}(\vec{x},\vec{p};t)=\mathcal{M}(\vec{\nabla}_{\vec{x}},\vec{p})\mathbbm{\vec{w}}(\vec{x},\vec{p};t) \ ,
\end{equation}
with $\mathbbm{\vec{w}}(\vec{x},\vec{p};t)$ and
$\mathcal{M}(\vec{\nabla}_{\vec{x}},\vec{p})$ satisfying
Eq.~(\ref{sec2a:dhw1}) -- (\ref{sec2a:dhw8}). We perform a derivative
expansion of the pseudo-differential operator $D_t$ in
Eq.~(\ref{sec2a:diffop_nonlocal}), such that:
\begin{equation}
\label{sec3a:dtexpansion}
D_t=\partial_t+eE(t)\partial_{p_3}+eE(t)\left[\Delta(x_3)\partial_{p_3}-\frac{\Delta^{\prime\prime}(x_3)}{24}\partial_{p_3}^3+...\right]
\end{equation}
where $\Delta^{\prime\prime}(x_3)$ denotes the second derivative with respect
to $x_3$. We may consider $\mathbbm{\vec{w}}(\vec{x},\vec{p};t)$ in an
expansion as well:
\begin{equation}
\vec{\mathbbm{w}}(\vec{x},\vec{p};t)=\mathbbm{\vec{w}}^{(0)}(\vec{p};t)+\mathbbm{\vec{w}}^{(1)}(\vec{x},\vec{p};t)
\ ,
\end{equation}
with $\mathbbm{\vec{w}}^{(0)}(\vec{p};t)$ being the exact result for the case
of a spatially homogeneous electric field and
$\mathbbm{\vec{w}}^{(1)}(\vec{x},\vec{p};t)$ denoting a small deviation from
the zeroth-order solution due to \colG{the} spatial inhomogeneity. Neglecting terms of
the order $\Delta(x_3)\mathbbm{\vec{w}}^{(1)}(\vec{x},\vec{p};t)$, we obtain:
\begin{eqnarray} 
&\left[\partial_t+eE(t) \partial_{p_3}-\mathcal{M}(\vec{\nabla}_{\vec{x}},\vec{p})\right]\mathbbm{\vec{w}}^{(1)}(\vec{x},\vec{p};t) \approx \qquad \qquad\nonumber \\
&-eE(t)\left[\Delta(x_3)\partial_{p_3}-\frac{\Delta^{\prime\prime}(x_3)}{24}\partial_{p_3}^3\right]\mathbbm{\vec{w}}^{(0)}(\vec{p};t) \ ,
\end{eqnarray}
such that the spatially homogeneous solution
$\mathbbm{\vec{w}}^{(0)}(\vec{p};t)$ acts as a source term for the spatially
inhomogeneous solution. In order to estimate the parameter regime for which
the omission of derivatives higher than linear might be a good approximation,
we consider a simple model:
\begin{equation} \Delta(x_3)=\delta_0\cos\left(\tfrac{x_3}{L}\right) \ ,
\end{equation}
with $L$ being the length scale of the spatial variation and $\delta_0\ll1$
being its amplitude. Introducing the dimensionless variable
\begin{equation} \lambda=mL 
\end{equation}
which measures the spatial variation in units of the Compton wave length, we
obtain:
\begin{eqnarray}
  \label{sec3a:dhw_approximation} 
 &\left[\partial_t+eE(t)\partial_{p_3}-\mathcal{M}(\vec{\nabla}_{\vec{x}},\vec{p})\right]\mathbbm{\vec{w}}^{(1)}(\vec{x},\vec{p};t) \approx \qquad \qquad \nonumber \\
&\quad-eE(t)\Delta(x_3)\left[\partial_{p_3}+\frac{m^2}{24\,\lambda^2}\partial_{p_3}^3\right]\mathbbm{\vec{w}}^{(0)}(\vec{p};t) \ .
\end{eqnarray}
This equation serves as the starting point for our analysis of the influence
of a small spatial inhomogeneity. In order to estimate the influence of higher
derivatives, we compare \colG{the terms occurring in the derivative expansion of the
pseudo-differential operator $D_t$ in Eq.~\eqref{sec3a:dtexpansion}}:
\begin{eqnarray}
   &\partial_{p_3}\mathbbm{\vec{w}}^{(0)}(\vec{p};t) \quad \mathrm{with} \quad
\frac{m^2}{24\,\lambda^2} \partial^3_{p_3}\mathbbm{\vec{w}}^{(0)}(\vec{p};t)& \ .
\end{eqnarray}
In fact, by means of this procedure we {do not} quantify the overall influence
of the small spatial inhomogeneity $\Delta(x_3)$ on the Schwinger effect; for
this we would really have to solve the PDE system
Eq.~(\ref{sec3a:dhw_pde_full}). However, by means of the derivative expansion
we might estimate a parameter region for which the higher derivatives do not
play an important role such that we could restrict ourselves to the solution
of the first-order PDE system:
\begin{equation}
  \label{sec3a:dhw_pde_linear}
\left[\partial_t+eE(t)\partial_{p_3}\right]\mathbbm{\vec{w}}(\vec{x},\vec{p};t)=\mathcal{M}(\vec{\nabla}_{\vec{x}},\vec{p})\mathbbm{\vec{w}}(\vec{x},\vec{p};t)
\ .
\end{equation}
It is known from the analysis of the Schwinger effect in spatially homogeneous
electric fields that the orthogonal momentum solely acts as an additional mass
term and does not change the qualitative behavior. Thus, for simplicity, we
will restrict ourselves in the following to $\vec{p}_\perp=0$, such that we
deal only with the following DHW functions:
\begin{alignat}{8}
  \label{sec3a:dhw_s} &\mathbbm{s}(\hat{p}_3;t)& \ = \ & - &
\frac{2}{\hat{\omega}(\hat{p}_3)}&\chi^{1}(\hat{p}_3;t)& \ - \ &
\frac{2\hat{p}_3}{\hat{\omega}(\hat{p}_3)}&\chi^{2}(\hat{p}_3;t)& \ , \\
  \label{sec3a:dhw_v3} &\mathbbm{v}_3(\hat{p}_3;t)& \ = \ & - &
\frac{2\hat{p}_3}{\hat{\omega}(\hat{p}_3)}&\chi^{1}(\hat{p}_3;t)& \ + \ &
\frac{2}{\hat{\omega}(\hat{p}_3)}&\chi^{2}(\hat{p}_3;t)& \ , \\
  \label{sec3a:dhw_t13} &\mathbbm{t}_{1,3}(\hat{p}_3;t)& \ = \ & &
2\,&\chi^{3}(\hat{p}_3;t)& \ ,
\end{alignat}
where we have introduced the dimensionless phase-space kinetic momentum
$\hat{p}_3=\frac{p_3}{m}$ and the dimensionless energy variable
$\hat{\omega}(\hat{p}_3)=\frac{\omega(p_3)}{m}$.

\subsection{Example 1: Constant electric field}
\label{sec3b}

We explicitly calculated the time-independent coefficients
$\chi^{i=\{1,2,3\}}(\vec{p})$ in App.~\ref{appb}, \colG{such that the DHW functions Eq.~(\ref{sec3a:dhw_s}) -- (\ref{sec3a:dhw_t13}) for $\vec{p}_\perp=0$
and in terms of the dimensionless variables $\hat{p}_3$ and $\hat{\omega}(\hat{p}_3)$ are easily calculable.}
%
%
In the end, we want to compare the first with the third derivative of the source term in Eq.~(\ref{sec3a:dhw_approximation}):
\colG{
\begin{eqnarray}
  \label{sec3b:comparison}
 &-m\epsilon\Delta(x_3)\left[\partial_{\hat{p}_3}+\frac{1}{24\,\lambda^2}\partial_{\hat{p}_3}^3\right]\mathbbm{\vec{w}}^{(0)}(\hat{p}_3)& \ .
\end{eqnarray}}
\noindent Therefore, we explicitly determine these derivatives according to Eq.~(\ref{sec2b:dhw_pde}), {yielding}:
\begin{alignat}{4}
  &\partial_{\hat{p}_3}\mathbbm{s}^{(0)}\ &=&\ \frac{2\hat{p}_3}{\epsilon}\mathbbm{t}^{(0)}_{1,3} \ , 
  \qquad  \qquad  \nonumber \\
  \label{sec3b:first_deriv}
  &\partial_{\hat{p}_3}\mathbbm{v}_3^{(0)}\ &=&\ -\frac{2}{\epsilon}\mathbbm{t}^{(0)}_{1,3} \ , 
  \qquad  \qquad  \\
  &\partial_{\hat{p}_3}\mathbbm{t}_{1,3}^{(0)}\ &=&\ -\frac{2\hat{p}_3}{\epsilon}\mathbbm{s}^{(0)}+\frac{2}{\epsilon}\mathbbm{v}_3^{(0)} \ . \nonumber
\end{alignat}
and
\begin{alignat}{4} 
  &\partial_{\hat{p}_3}^3\mathbbm{s}^{(0)}\ &=&\ 
  \frac{4}{\epsilon^2}\left[-3\hat{p}_3\mathbbm{s}^{(0)}+2\mathbbm{v}_3^{(0)}-\tfrac{2\hat{p}_3\hat{\omega}^2(\hat{p}_3)}{\epsilon}\mathbbm{t}_{1,3}^{(0)}\right]\ ,  \nonumber \\
  \label{sec3b:third_deriv}
  &\partial_{\hat{p}_3}^3\mathbbm{v}_3^{(0)}\ &=&\ 
  \tfrac{4}{\epsilon^2}\left[\mathbbm{s}^{(0)}+\tfrac{2\hat\omega^2(\hat{p}_3)}{\epsilon}\mathbbm{t}_{1,3}^{(0)}\right] \ ,\qquad   \ 
  \\
  &\partial_{\hat{p}_3}^3\mathbbm{t}_{1,3}^{(0)}\ &=&\ 
  \tfrac{4}{\epsilon^2}\left[\tfrac{2\hat{p}_3\hat\omega^2(\hat{p}_3)}{\epsilon}\mathbbm{s}^{(0)}-\tfrac{2\hat\omega^2(\hat{p}_3)}{\epsilon}\mathbbm{v}_3^{(0)}-3\hat{p}_3\mathbbm{t}_{1,3}^{(0)}\right] \ ,  \nonumber
\end{alignat}
where we have dropped the arguments of the DHW functions for simplicity. 

\begin{figure}[b]
  \subfigure{\includegraphics[width=6.5cm]{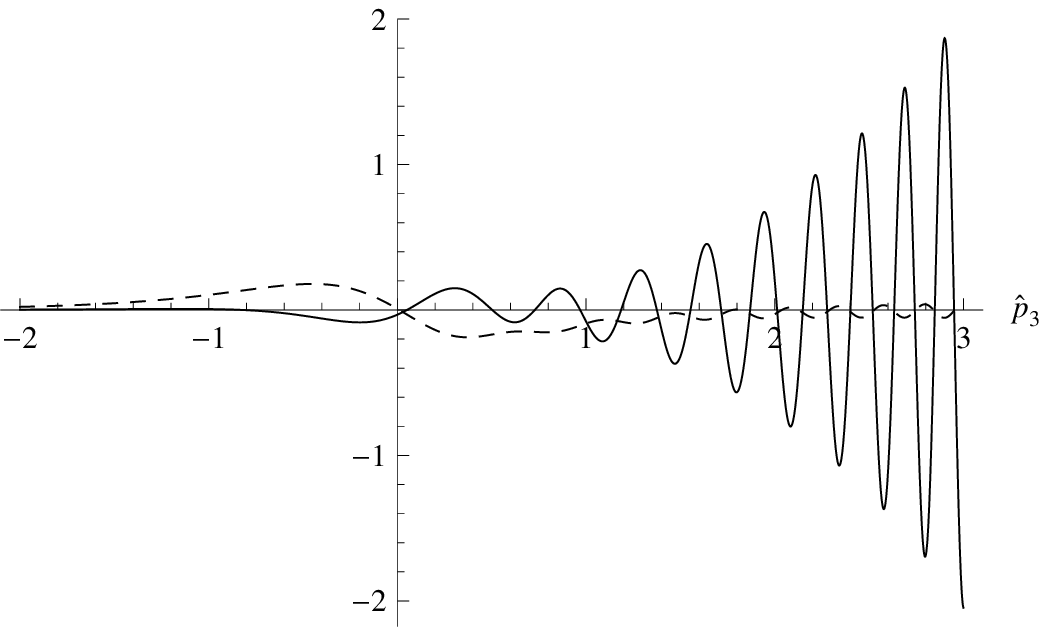}}
  \hspace{1cm}
  \subfigure{\includegraphics[width=6.5cm]{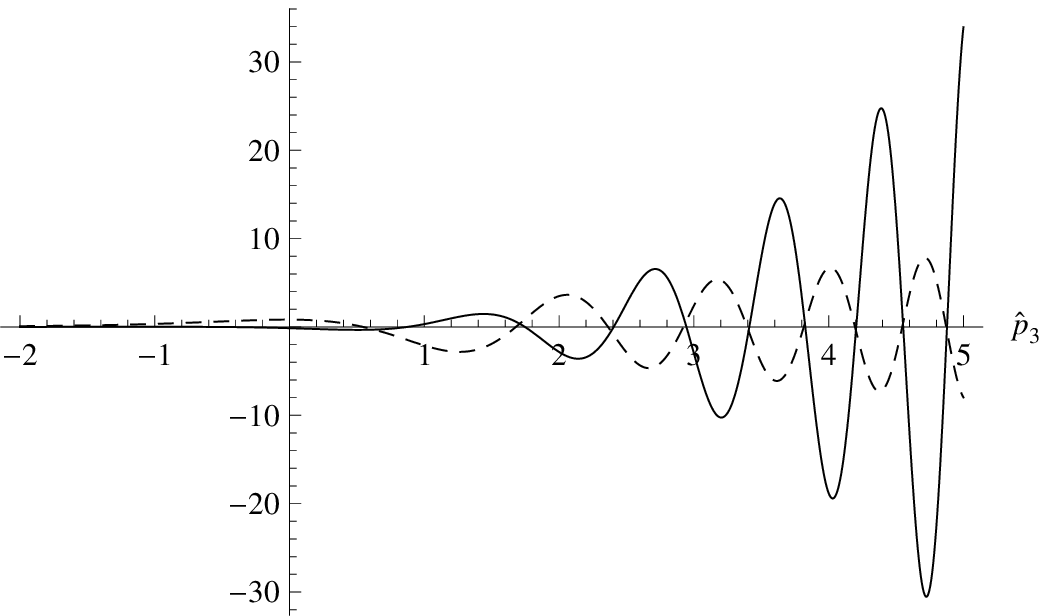}}
  \caption{\label{sec3b:fig_comparison} Comparison of the leading-order
    derivative term $\partial_{\hat{p}_3}\mathbbm{t}_{1,3}^{(0)}(\hat{p}_3)$
    (dashed) with the next-to-leading order term $\frac{1}{24}\partial_{\hat{p}_3}^3\mathbbm{t}_{1,3}^{(0)}(\hat{p}_3)$
    (solid) for $E(t)=E_0$ with $\epsilon=E_0/\Ec= 0.2$ (\colG{upper}) and $\epsilon=1$
    (\colG{lower}). For higher momenta, the next-to-leading order term eventually
    exceeds the leading-order term as the electric field persistently
    accelerates the produced pairs.}
\end{figure}

According to Eq.~(\ref{sec3b:comparison}){,} we compare
$\partial_{\hat{p}_3}\mathbbm{\vec{w}}^{(0)}(\hat{p}_3)$ with
$\frac{1}{24} \partial_{\hat{p}_3}^3\mathbbm{\vec{w}}^{(0)}(\hat{p}_3)$ in the
following; due to the fact that all the DHW functions show a rather similar
behavior, we restrict ourselves to $\mathbbm{t}_{1,3}^{(0)}(\hat{p}_3)$ for
simplicity. Fig.~\ref{sec3b:fig_comparison} clearly shows that the higher
derivatives always become more important than the first derivative for large
momenta $\hat{p}_3$. This might be understood in the following way: Due to the
acceleration in the electric field, all length scales are ultimately probed and
become important, even though the pair creation process happens on the length
scale of \colG{$\mathcal{O}(\lambda)$}. Note, however, that the point at which the higher
derivatives are of the order of the first derivative depends on the electric
field strength $\epsilon=E_0/\Ec$: For higher field strengths this point is
already reached for lower momenta.

\begin{figure}[t]
 \subfigure{\includegraphics[width=6.5cm]{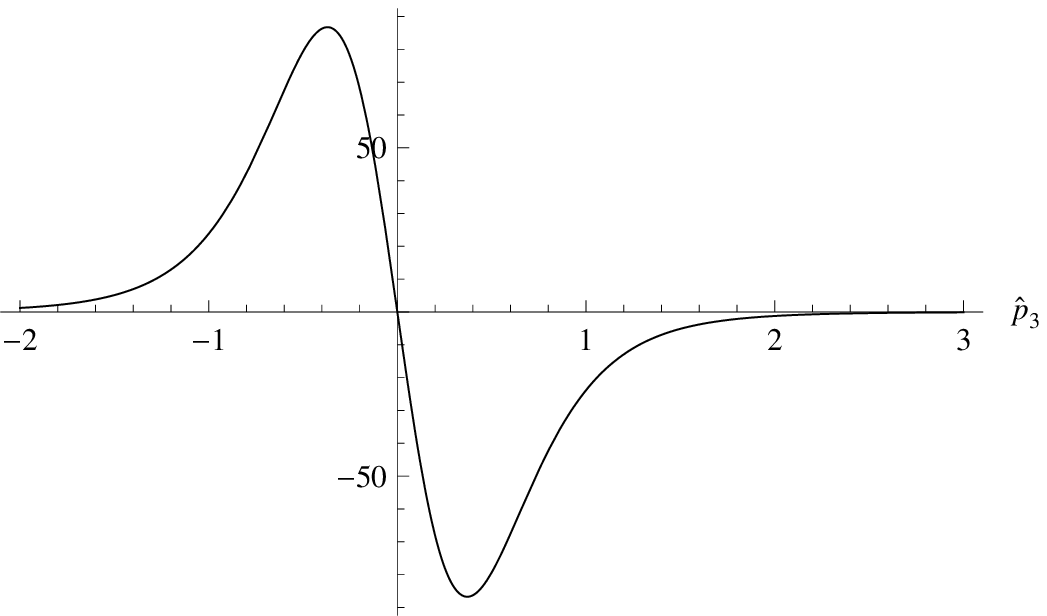}}
 \hspace{1cm}
 \subfigure{\includegraphics[width=6.5cm]{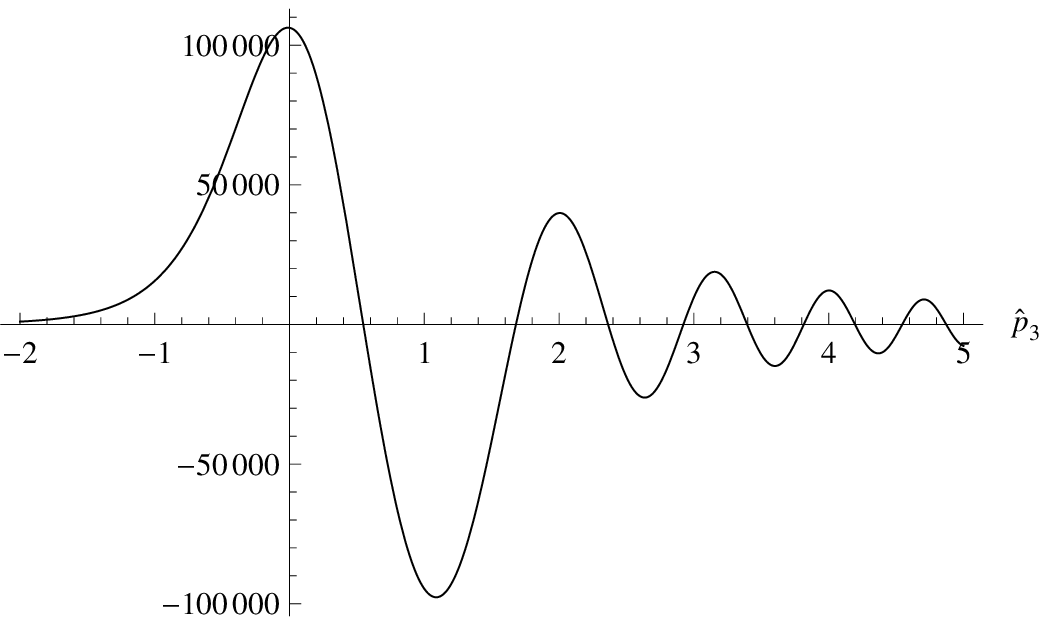}}
 \caption{\label{sec3b:fig_rate} Pair production rate for $E(t)=E_0$ with
   $\epsilon=E_0/\Ec= 0.2$ (\colG{upper}) and $\epsilon=1$ (\colG{lower}). For
   $\epsilon=0.2$, pair production occurs dominantly for momenta
   $p_3\lesssim2$ where next-to-leading order derivative terms remain small,
   cf. Fig.~\ref{sec3b:fig_comparison}. By contrast, higher derivative terms
   are expected to take a quantitative influence on pair production for
   $\epsilon=1$ and sizable spatial variation $\lambda$. }
\end{figure}
 
In order to estimate the importance of the spatial inhomogeneity for the pair
creation process itself, we should have a closer look at the pair production
rate in Fig.~\ref{sec3b:fig_rate}:

For $\epsilon=0.2$, the dominant contributions to the pair production rate
stem from a region of kinetic momenta up to $\hat{p}_3\approx\pm 2$. In this
regime, the first derivatives are still of the order of the third derivatives,
as shown in Fig.~\ref{sec3b:fig_comparison}. As a consequence, the effect of
higher derivatives on the pair production process should be taken into account
only for a spatial variation of the Compton
wavelength $\lambda=\mathcal{O}(1)$, whereas the effect of the higher derivatives becomes suppressed for
larger \colG{variation scales $\lambda$} . 

\begin{figure}[b]
 \subfigure{\includegraphics[width=6.5cm]{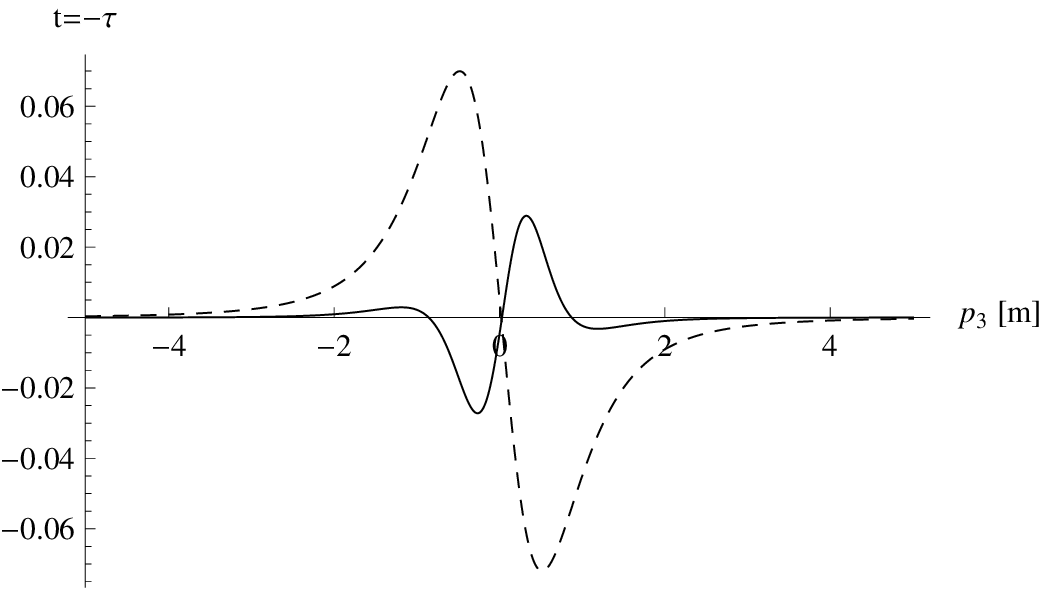}}
 \hspace{0.5cm}
 \subfigure{\includegraphics[width=6.5cm]{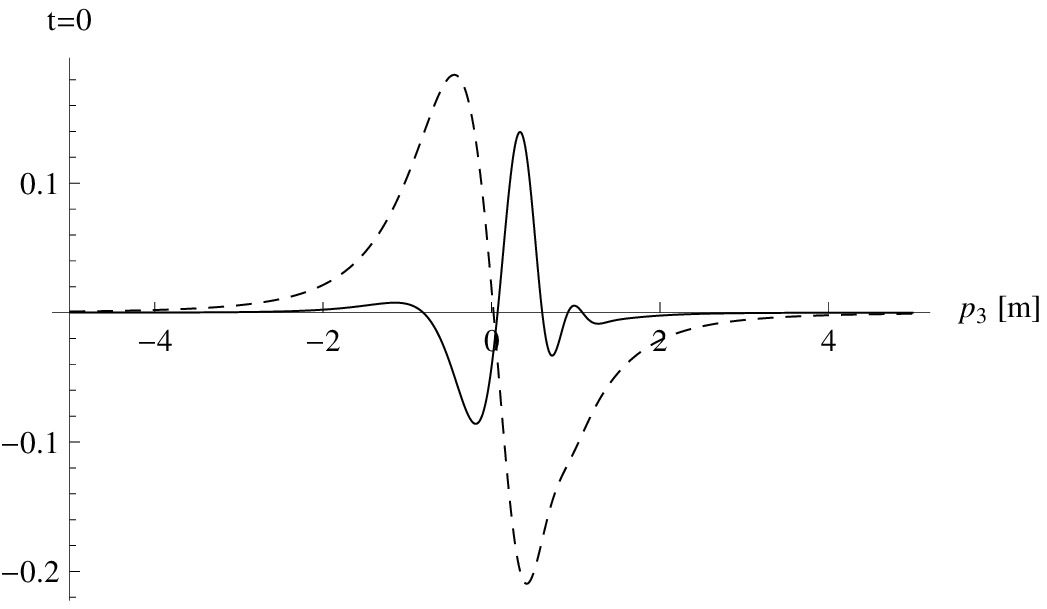}}
 \hspace{0.5cm}
 \subfigure{\includegraphics[width=6.5cm]{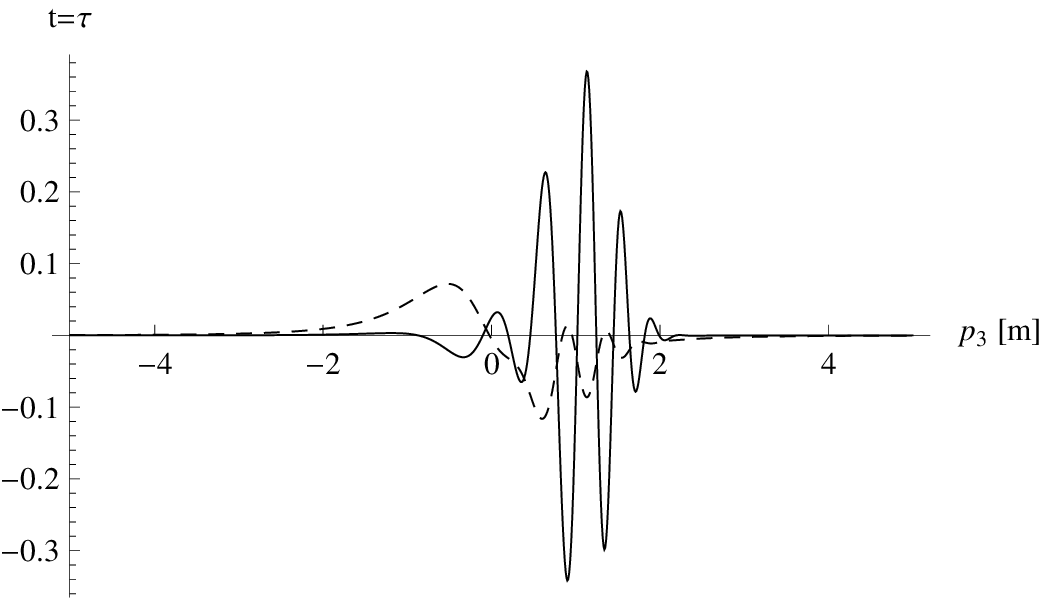}}
 \caption{\label{sec3c:fig_e02g1} Comparison of the leading-order derivative
   term $\partial_{\hat{p}_3}\mathbbm{t}_{1,3}^{(0)}(\hat{p}_3;u)$ (dashed)
   with the next-to-leading order term
   $1/24\partial_{\hat{p}_3}^3\mathbbm{t}_{1,3}^{(0)}(\hat{p}_3;u)$ (solid)
   for $\epsilon=E_0/\Ec =0.2$ and $\gamma=1$ for $t=-\tau$ (upper), $t=0$ (middle) and
   $t=\tau$ (lower).}
\end{figure}

For $\epsilon=1$, the situation is slightly different: Again, the dominant
contributions to the pair production rate arise from kinetic momenta
up to $\hat{p}_3 \approx\pm2$, however, there are non-vanishing contributions
for higher momenta as well, as shown in Fig.~\ref{sec3b:fig_comparison}. Due
to the fact that the higher derivatives become more important than the first
derivatives for large momenta, the scale of spatial variation $\lambda$ has to
increase as well in order to suppress the higher derivatives.

To conclude: Concerning the pair production process, there is a strong
interplay between the electric field strength $\epsilon$ and the scale of
spatial variation $\lambda$. This interplay between field strength and scale
of spatial variation affecting the quality of the derivative expansion has
already been observed in earlier studies \cite{Gies:2005bz}. In order to keep
the pair production process itself unaltered by higher derivatives, the scale
of spatial variation $\lambda$ must get larger for higher electric field
strengths $\epsilon$. However, even if the effect of the higher derivatives on
the pair production process might be negligible, the effect on the final
momentum distribution might be large: this is due to the acceleration of the
pairs in the electric field which finally emphasizes higher momenta such that
higher derivatives always become more important than the first derivative.

\subsection{Example 2: Sauter-type electric field}
\label{sec3c}

For the Sauter-type electric field we may perform a similar analysis like for
the constant electric field. Again, we only consider $\vec{p}_\perp=0$,
however, there is a huge qualitative difference since the DHW functions now
depend on both the phase-space kinetic momentum $\hat{p}_3$ and the time
variable $u$.  Using the expressions given in App.~\ref{appb}, we are able to
analytically calculate the DHW functions Eq.~(\ref{sec3a:dhw_s}) --
(\ref{sec3a:dhw_t13}). Again, we consider the source term in
Eq.~(\ref{sec3a:dhw_approximation}), which now reads:
\colG{
\begin{eqnarray}
  \label{sec3c:comparison}
 -4m\epsilon u(1-u)\Delta(x_3)\left[\partial_{\hat{p}_3}+\tfrac{1}{24\,\lambda^2}\partial_{\hat{p}_3}^3\right]\mathbbm{\vec{w}}^{(0)}(\hat{p}_3;u) \ .
\end{eqnarray}}
\noindent In what follows, we compare the first derivative $\partial_{\hat{p}_3}\mathbbm{\vec{w}}^{(0)}(\hat{p}_3;u)$ with the third derivatives $\frac{1}{24} \partial_{\hat{p}_3}^3\mathbbm{\vec{w}}^{(0)}(\hat{p}_3;u)$. Again, we restrict ourselves to $\mathbbm{t}_{1,3}^{(0)}(\hat{p}_3;u)$. Nonetheless, there are three big differences in comparison to the constant electric field: First, we are not able to calculate the first and third derivative with respect to $\hat{p}_3$ as easily as for the constant electric field, as shown in Eq.~(\ref{sec3b:first_deriv}) -- (\ref{sec3b:third_deriv}), since these derivatives now include parameter derivatives of the Gauss hypergeometric \colG{function}. Nevertheless, a numerical calculation is rather simple. Second, the situation is not quasi-static anymore but it makes a difference at which moment of time the system is considered. Third, the field strength $\epsilon$ is not the only relevant parameter but we also have to consider the dependence on the pulse length $\tau$ via the Keldysh parameter $\gamma=1/(m\epsilon\tau)$. In order to analyze the interplay between these quantities, we consider the system at three different instants of time: At $t=-\tau$ (increase of field strength), $t=0$ (maximum of field strength) and $t=\tau$ (decrease of field strength).

\begin{figure}[b]
 \subfigure{\includegraphics[width=6.5cm]{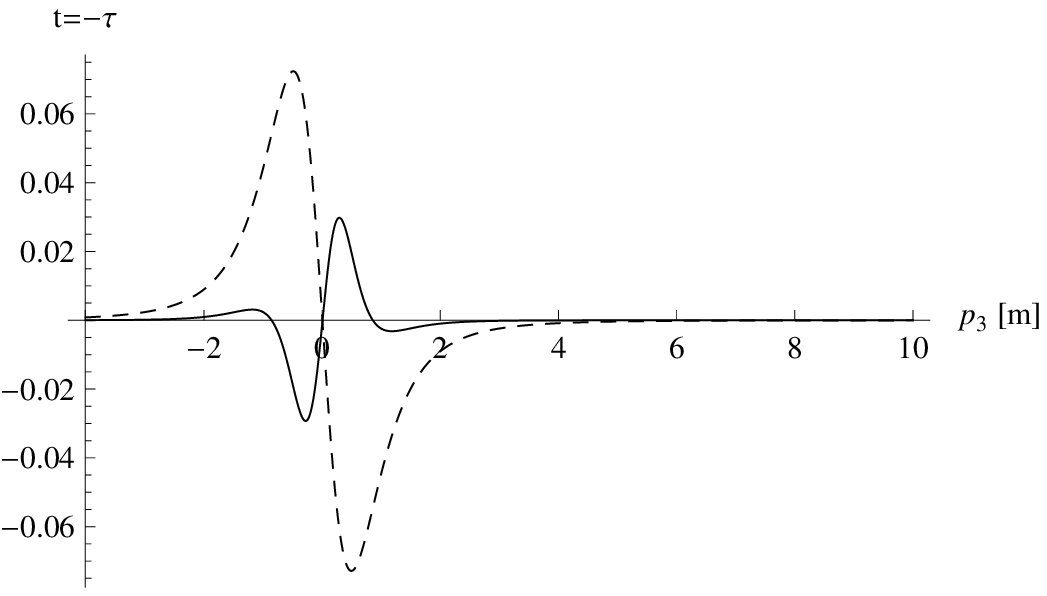}}
 \hspace{0.5cm}
 \subfigure{\includegraphics[width=6.5cm]{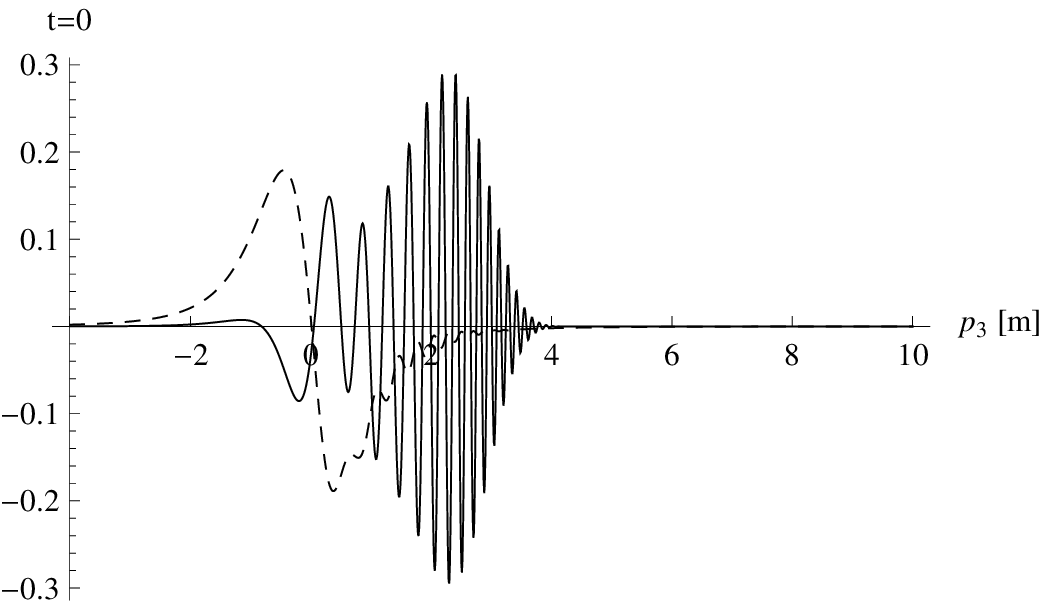}}
 \hspace{0.5cm}
 \subfigure{\includegraphics[width=6.5cm]{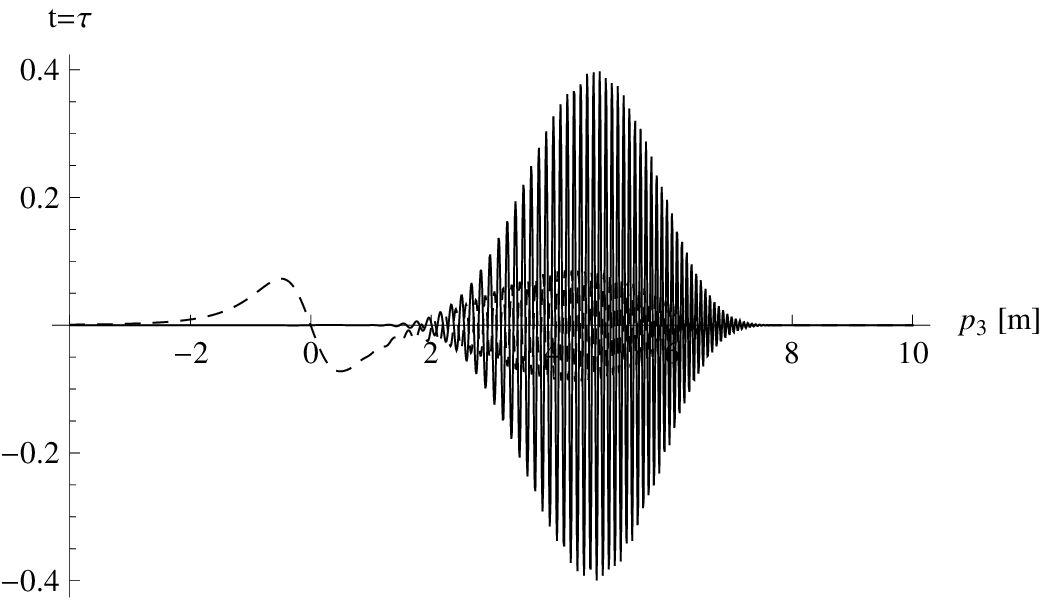}}
 \caption{\label{sec3c:fig_e02g02} Comparison of the leading-order derivative
   term $\partial_{\hat{p}_3}\mathbbm{t}_{1,3}^{(0)}(\hat{p}_3;u)$ (dashed)
   with the next-to-leading order term
   $1/24\partial_{\hat{p}_3}^3\mathbbm{t}_{1,3}^{(0)}(\hat{p}_3;u)$ (solid)
   with $\epsilon=E_0/\Ec =0.2$ and $\gamma=0.2$ for $t=-\tau$ (\colG{upper}), $t=0$ (\colG{middle})
   and $t=\tau$ (\colG{lower}). Note: For $t=\tau$ the third derivative is in fact 2
   magnitudes larger than the first derivative but has been scaled by a factor
   of $0.01$.}
\end{figure}

Let us first concentrate on Fig.~\ref{sec3c:fig_e02g1} with $\epsilon=0.2$ and
$\gamma=1$, where we already observe some general features: As in the case of
the constant electric field, the third derivatives show an out-of-phase
behavior compared to the first derivative: At momentum values where the third
derivatives show a local maximum, the first derivative shows a local minimum
and vice versa. As a consequence, depending on the scale of spatial variation
$\lambda$, the influence of 
the first derivative might be inverted due to the third derivative term. We also observe that the
relative importance of the third derivatives in comparison to the first
derivative becomes bigger for later times. This might be interpreted in the
following way: At late times the created particles have had more time to be
accelerated in the electric field and, as a consequence, have travelled over
larger distances and were exposed to even large scale inhomogeneities. The
effect of acceleration in the electric field might also be seen in the shift
of the global maximum of the third derivatives towards  higher momentum values. If
we want to suppress the higher derivatives in comparison to the first
derivative, we should choose the scale of spatial variation to be at least of
the order $\lambda\gtrsim\mathcal{O}(5)$.

Next we switch to a longer pulse with $\gamma\to0.2$ as shown in
Fig.~\ref{sec3c:fig_e02g02}, which corresponds to a pulse with the same
electric field strength but with a 5-times longer duration. We see that the
behavior at early times $t=-\tau$ is very similar, however,  the
situation changes drastically for later times: First, due to the longer pulse
duration the created particles are accelerated to higher
momenta. Additionally, we observe a strong enhancement of the oscillatory
structure and a strong increase of the magnitude of the third
derivatives. Especially at $t=\tau$, the third derivatives are orders of
magnitude larger than the first derivatives. This means that the scale of
spatial variation $\lambda$ has to be chosen much larger in order to suppress
the influence of higher order derivatives.

The change in the overall magnitude is the distinctive feature when we
increase the field strength $\epsilon\to1$ as shown in
Fig.~\ref{sec3c:fig_e1g02}. The general behavior is rather similar as before,
however, the source term becomes more important in comparison to the left-hand
side of Eq.~(\ref{sec3a:dhw_approximation}). As a consequence, we expect the
overall effect of spatial inhomogeneities to be more important for strong
electric fields than for weak electric fields.

\begin{figure}[t]
 \subfigure{\includegraphics[width=6.5cm]{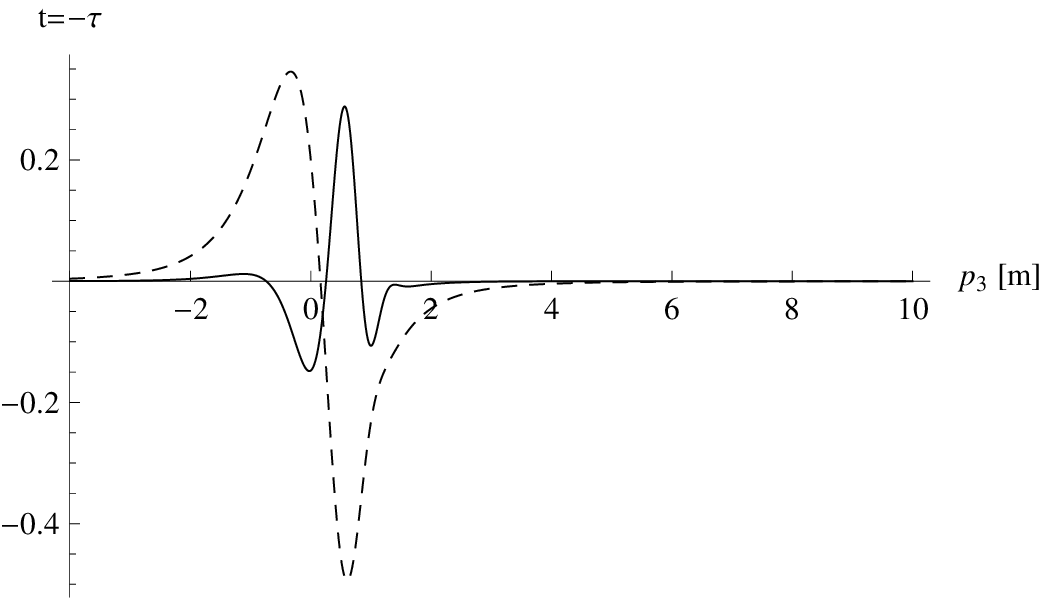}}
 \hspace{0.5cm}
 \subfigure{\includegraphics[width=6.5cm]{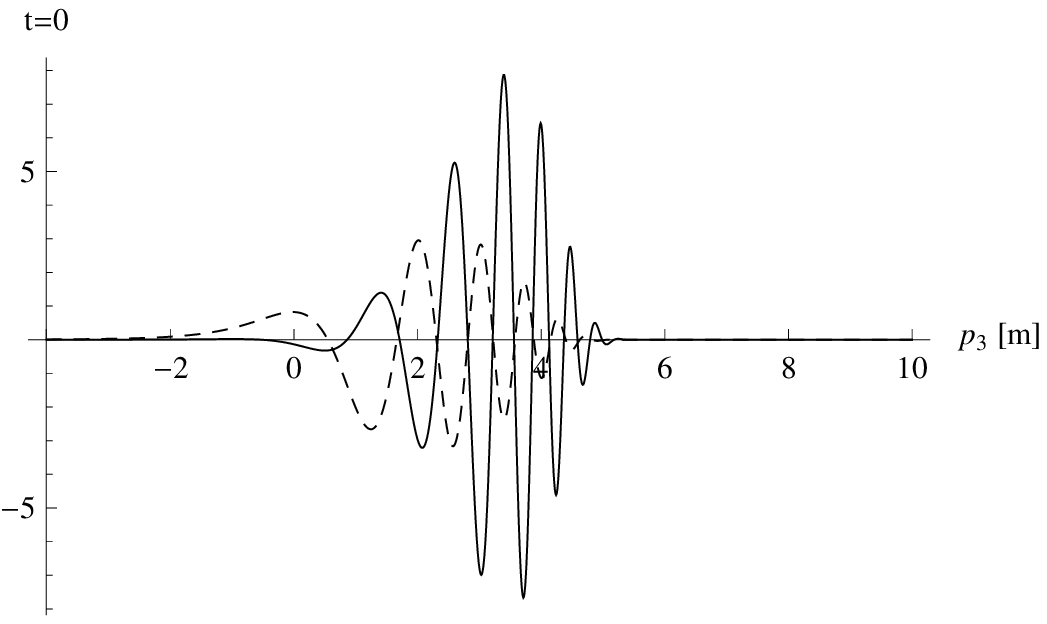}}
 \hspace{0.5cm}
 \subfigure{\includegraphics[width=6.5cm]{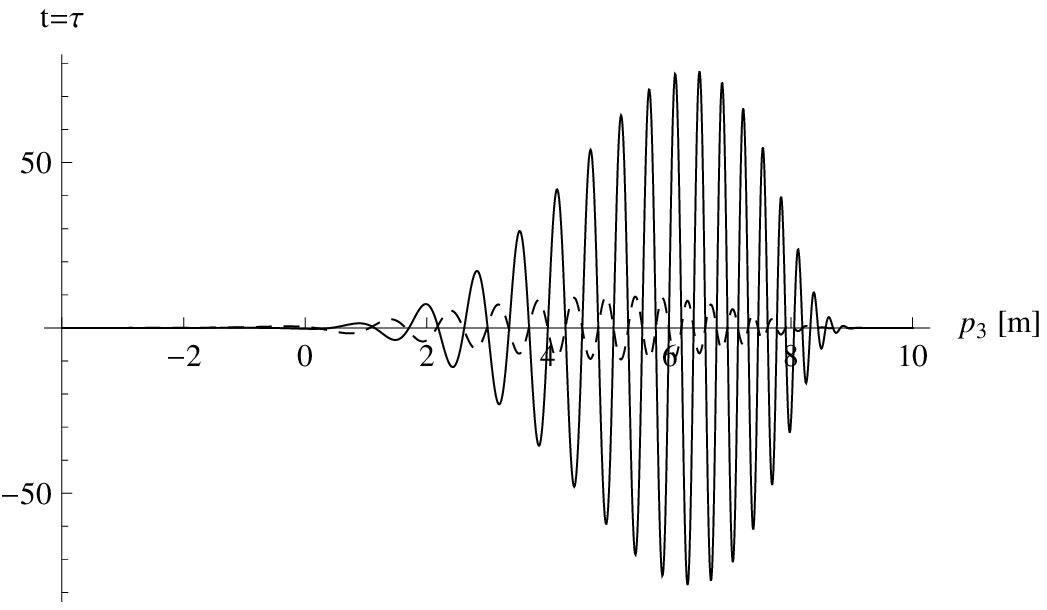}}
 \caption{\label{sec3c:fig_e1g02} Comparison of the leading-order derivative
   term $\partial_{\hat{p}_3}\mathbbm{t}_{1,3}^{(0)}(\hat{p}_3;u)$ (dashed) with the next-to-leading order term
   $1/24\partial_{\hat{p}_3}^3\mathbbm{t}_{1,3}^{(0)}(\hat{p}_3;u)$ (solid)
   with $\epsilon=1$ and $\gamma=0.2$ for $t=-\tau$ (\colG{upper}), $t=0$ (\colG{middle}) and
   $t=\tau$ (\colG{lower}).} 
\end{figure}

To summarize, Fig.~\ref{sec3c:fig_e02g1} -- \ref{sec3c:fig_e1g02} give us the
following physical picture:

\begin{enumerate} 
\item[(a)]{Concerning the time variable $u$, we see that the role of higher
    derivatives is more important at \colG{late} times than at early times. This
    suggests that even if the influence of spatial inhomogeneities on the pair
    production process itself might not be very strong, the final momentum
    distribution can be strongly altered.}
\item[(b)]{Concerning the field strength $\epsilon$, we conclude by comparing
    the magnitude of the source term for the different values of $\epsilon$
    that the influence of spatial inhomogeneities on the pair production
    process becomes more important for higher field strengths. We interpret
    this observation as arising from the fact that for weak fields the created
    particles are not accelerated that much and, as a consequence, they do
    only \colG{feel} inhomogeneities on large length scales but not on shorter
    variation scales.}
  \item[(c)]{Concerning the Keldysh parameter $\gamma$ we note that small
      values of $\gamma$ correspond either to strong electric field strengths
      or to longer pulse durations $\tau$. Therefore, particles are more
      substantially accelerated in the electric field and, as a consequence,
      spatial inhomogeneities might have a greater importance, since the
      particles feel the inhomogeneities even on shorter scales of variation.}
\end{enumerate}

As a consequence, we observe that there will generally be a complex interplay
between all the relevant parameters for any type of space- and time-dependent
electric field $\vec{E}(\vec{x},t)$. Thus, it is difficult to predict {\it{a
    priori}} whether neglecting higher derivatives can be a good approximation
or not. It is clear that in the limit of a spatially homogeneous electric
fields $\lambda\to\infty$ the leading-order derivative approximation
Eq.~(\ref{sec3a:dhw_pde_linear}) becomes exact since all higher derivatives
are suppressed by a factor of $\lambda^{-2n}$. However, it is not clear {\it{a
    priori}} for which values of $\lambda$ the higher derivatives play a
quantitatively important role and should be taken into account. Future
investigations on that problem should help to better understand the interplay
between the different scales.

\section{Conclusions \& Outlook}
\label{sec4}

We have investigated the Dirac-Heisenberg-Wigner (DHW) formalism for
non-perturbative pair production in general electromagnetic fields. As a
genuine real-time formalism, this approach provides for a comprehensive
framework of addressing all aspects of pair production most notably, the
non-equilibrium character of pair-production in a fully \colG{time-} and \colG{space-resolved} manner. 

We have shown that the DHW formalism includes quantum kinetic theory (QKT)
which has so far been the most successful approach to describe the real-time
evolution of pair production in the limit of time-dependent but spatially
homogeneous electric fields. We conclude that the DHW formalism provides for
the desired generalization of QKT to the case of arbitrarily general space- and
time-dependent electromagnetic fields. For a given field, the solution of the
DHW formalism is parameterized in spinor QED in four-dimensional spacetime by
16 irreducible components of the Wigner function which encode the phase-space
distributions of physical quantities such as mass, charge and current
densities as provided by the produced pairs. From the knowledge of these
quantities, \colG{although they are no semi-positive definite probability distributions,} 
physical observables such as the pair distribution function in
phase space can directly be inferred. 

Whereas the DHW formalism is completely general as far as the details of the
external field are concerned, we have confined ourselves in the present work
to an analysis of exactly solvable cases such as the constant electric field
and the Sauter potential. Of course, such exactly solvable cases always
provide for a controlled starting point for more general cases, in particular,
they should also serve as a benchmark for future full numerical
studies. Moreover, we have used these cases in the present work to provide for
a first glance at the possible use and limitations of natural approximation
schemes such as the derivative expansion. The leading-order of this
approximation corresponds to a {\em locally-constant field approximation},
i.e., approximating the spatial dependence of the field locally by a constant
field.

The picture arising from this investigation is rather diverse: it has already
been known, for instance, from worldline instanton studies
\cite{Dunne:2005sx,Dunne:2006st}, that the locally-constant field
approximation underestimates the pair-production rate for fields varying in
time, and overestimates the rate for fields varying in space. Whereas the
exact solution, e.g. for the Sauter potential in Sect.~\ref{sec:sauter}
reflects this fact, we observe that general statements about the potential
quality of the derivative or locally-constant-field approximation cannot
straightforwardly be made. For all concrete examples, we observe that the
next-to-leading order derivative terms in fact exceed the leading-order terms
either for \colG{higher momenta} or for late times. Taken at face value,
this seems to imply that the derivative expansion is always bound to fail as
soon as the field exhibits spatial variations. However, the reason for this
strong modification of the next-to-leading order terms lies in the fact that
the persistent presence of accelerating field components, of course, exerts a
strong influence on high-momentum components which eventually resolve also
small spatial variations of the field. Therefore, it is only natural to expect
that the derivative expansion should fail in the way it does for high momentum
components and at late times.

Nevertheless, our results also provide a guideline to a less strict view on
the quality of the derivative expansion: phenomenologically, the most relevant
quantity is the pair distribution function in momentum space. Quantitatively,
the question needs to be addressed whether the next-to-leading order terms of
the derivative expansion exert a strong influence on this distribution
function. As the higher derivative terms become dominant for higher-momentum
components, we conclude that the derivative expansion can still remain a
reasonable approximation as long as the dominant pair distribution is peaked
at lower momenta. Whether or not this is the case, depends not only on the
scale of spatial variation of the field, but also on the overall field
strength and also possible further time dependencies. As a rule of thumb, we
observe that the dominant low-momentum components appear to remain little
affected at \colG{significantly} subcritical field strengths with spatial variations being \colG{substantially} larger
than the Compton wavelength.

Beyond the technical question about the quality of the derivative expansion,
the most interesting question is certainly as to whether a phenomenologically
relevant interplay between characteristic signatures of pair production and
space-time shaping of the external field \colG{exists}. Based on the suprising
observations that have already been made for simple time-dependencies
\cite{Hebenstreit:2009km,Dumlu:2010vv}, we expect that this question can be
answered in the affirmative. From our present studies of spatial variations
and the dominant effect on high-momentum components, we conclude that temporal
and spatial pulse shaping can have substantial effects on the momentum
structure of the pair distribution function. As to whether temporal and
spatial pulse shaping can also optimize the total number of produced pairs
certainly remains the most pressing question which we hope to address with the
DHW formalism  based on full numerical solutions in the near future. 

\begin{acknowledgments}
We are grateful to Gerald V.~Dunne for helpful discussions. \colG{This work is supported by the DOC program of the Austrian Academy of Sciences, by the FWF doctoral program DK-W1203 (FH) and by the DFG through grants SFB/TR18, GRK1523, and Gi 328/5-1 (HG).}
\end{acknowledgments}

\begin{appendix}
\section{Quantum Kinetic Theory (QKT)}
\label{appa}

In this appendix, we give a brief derivation of the quantum kinetic equation
describing Schwinger pair production in spatially homogeneous, time-dependent
electric fields \cite{Smolyansky:1997fc, Schmidt:1998vi}. As in the main part
of this paper, we adopt the temporal gauge $A_0=0$. We choose the vector
potential $\vec{A}(t)=A(t)\vec{e}_3$ such that the Dirac equation reads:
\begin{equation}
  \left(i\gamma^0\partial_t+i\vec{\gamma}\cdot\left[\vec{\nabla}_{\vec{x}}-ieA(t)\vec{e}_3\right]-m\right)\Psi(\vec{x},t)=0. 
\end{equation}
Due to spatial homogeneity, we decompose the spinor field into its Fourier
modes according to Eq.~(\ref{sec2b:fourier_decomp}), such that the Dirac
equation for the mode function $\widetilde{\psi}(\vec{q},t)$ reads:
\begin{equation}
  \left(i\gamma^0\partial_t-\vec{\gamma}\cdot\vec{\pi}(\vec{q},t)-m\right)\widetilde{\psi}(\vec{q},t)=0 \ .
\end{equation}
Again note that $\vec{\pi}(\vec{q},t)=\vec{q}-eA(t)\vec{e}_3$ denotes the
time-dependent kinetic momentum on the trajectory whereas $\vec{q}$ denotes
the canonical momentum. In order to solve this equation we apply the ansatz:
\begin{equation}
  \widetilde{\psi}(\vec{q},t)=\left(i\gamma^0\partial_t-\vec{\gamma}\cdot\vec{\pi}(\vec{q},t)+m\right)\widetilde{\phi}(\vec{q},t)
  \ ,
\end{equation}
such that the spinor-valued function $\widetilde{\phi}(\vec{q},t)$ obeys the equation:
\begin{equation}
  \label{appa:eom_phi}
  \left(\partial_t^2+\widetilde{\omega}^2(\vec{q},t)+ieE(t)\gamma^0\gamma^3\right)\widetilde{\phi}(\vec{q},t)=0 \ , 
\end{equation}
with $\widetilde{\omega}(\vec{q},t)$ defined as before. It is convenient to expand $\widetilde{\phi}(\vec{q},t)$ in a basis consisting of the eigenvectors of $\gamma^0\gamma^3$, such that:
\begin{equation}
  \label{appa:ansatz}
  \widetilde{\phi}(\vec{q},t)=\sum_{s}R_s\widetilde{g_s}(\vec{q},t) \quad \mathrm{with} \quad \gamma^0\gamma^3R_s=\lambda R_s \ .
\end{equation}
There are two eigenvectors $R_{s=1,2}$ with $\lambda=+1$ and two eigenvectors $R_{s=3,4}$ with $\lambda=-1$. Inserting this ansatz into Eq.~(\ref{appa:eom_phi}), each $\widetilde{g_s}(\vec{q},t)$ obeys the equation of a time-dependent oscillator:
\begin{equation}
  \label{appa:eom_g}
  \left(\partial_t^2+\widetilde{\omega}^2(\vec{q},t)+ieE(t)\lambda\right)\widetilde{g_s}(\vec{q},t)=0 \ , \\
\end{equation}
which are in general not exactly solvable; exceptions are the constant electric field $E(t)=E_0$ and the Sauter-type electric field $E(t)=E_0\operatorname{sech}^2(t/\tau)$ (see Sec. \ref{sec2c}). Each of them is a second-order differential equation and possesses as such two linearly independent solutions $\widetilde{g_s}^{\pm}(\vec{q},t)$. Due to the fact that Eq.~(\ref{appa:eom_phi}) gives four equations but the ansatz Eq.~(\ref{appa:ansatz}) allows for eight solutions, there is a redundancy which is removed by choosing only one set of eigenvectors, either $s=\{1,2\}$ or $s=\{3,4\}$. Due to the absence of magnetic fields, we impose the same initial conditions for both spin states, such that:
\begin{equation}
  \widetilde{g_1}^{(\pm)}(\vec{q},t)=\widetilde{g_2}^{(\pm)}(\vec{q},t)=\widetilde{g}^{(\pm)}(\vec{q},t) \ .
\end{equation}
Consequently, we canonically quantize $\widetilde{\psi}(\vec{q},t)$ according to Eq.~(\ref{sec2b:quantization}) by introducing anti-commuting creation/annihilation operators as well as four-spinors, with:
\begin{alignat}{3}
  \label{appa:spinoru1}
  \widetilde{u_s}(\vec{q},t)&\ = \ &\left(i\gamma^0\partial_t-\vec{\gamma}\cdot\vec{\pi}(\vec{q},t)+m\right)\widetilde{g}^{(+)}(\vec{q},t)R_s& \ , \\
  \label{appa:spinorv1}
  \widetilde{v_s}(-\vec{q},t)&\ = \ &\left(i\gamma^0\partial_t-\vec{\gamma}\cdot\vec{\pi}(\vec{q},t)+m\right)\widetilde{g}^{(-)}(\vec{q},t)R_s& \ .
\end{alignat}
Due to the fact that we work in the Heisenberg picture, the
creation/annihilation operators are time-dependent in general, however, due to
the choice Eq.~(\ref{appa:spinoru1}) -- (\ref{appa:spinorv1}), the whole
time-dependence can be absorbed into the four-spinors. It can be shown, that
in the case of vanishing electric fields, the properly normalized vacuum
solutions are given by:
\begin{equation}
  \label{appa:g_free}
  \widetilde{g}^{(\pm)}_{\mathrm{vac}}(\vec{q},t)=\frac{e^{\mp i\omega(\vec{q})t}}{\sqrt{2\omega(\vec{q})(\omega(\vec{q})\mp q_3)}}  \ ,
\end{equation}
with $\omega(\vec{q})=\sqrt{m^2+\vec{q}^2}$. It is important to note that a
particle/antiparticle interpretation of the field quanta is only possible in
the case of such plane-wave solutions. However, as soon as electric fields are
present, the mode functions $\widetilde{g}^{(\pm)}(\vec{q},t)$ are no plane
waves anymore and an interpretation in terms of particles/antiparticles is not
straightforward. It is a further consequence of the presence of electric
fields that the Hamiltonian operator achieves off-diagonal elements which
account for particle/antiparticle creation/annihilation.

The Hamiltonian operator might be diagonalized by performing a unitary non-equivalent change of basis to a quasi-particle representation via a time-dependent Bogoliubov transformation:
\begin{alignat}{4}
  \label{app:bogoliubov1}
  &\widetilde{a}_s(\vec{q},t)& \ = \ &\widetilde{\alpha}(\vec{q},t)a_s(\vec{q})-\widetilde{\beta}^*(\vec{q},t)b_s^\dagger(-\vec{q}) \ ,
  \\ 
  \label{app:bogoliubov2}
  &\widetilde{b}_s^\dagger(-\vec{q},t)& \ = \ &\widetilde{\beta}(\vec{q},t)a_s(\vec{q})+\widetilde{\alpha}^*(\vec{q},t)b_s^\dagger(-\vec{q})
  \ ,
\end{alignat}
with the creation/annihilation operators becoming time-dependent but still fulfilling the equal-time anticommutation relations. In order to be a canonical transformation, the Bogoliubov coefficients $\widetilde{\alpha}(\vec{q},t)$ and $\widetilde{\beta}(\vec{q},t)$ have to fulfill:
\begin{equation}
  |\widetilde{\alpha}(\vec{q},t)|^2+ |\widetilde{\beta}(\vec{q},t)|^2=1 \ .
\end{equation}
In pure vacuum when no electric fields are present, the two different operator bases coincide such that $\widetilde{\alpha}_\mathrm{vac}(\vec{q},t)=1$ and $\widetilde{\beta}_\mathrm{vac}(\vec{q},t)=0$. Note, that this relation also holds in the presence of electric fields at asymptotic times $t\to-\infty$.  Within this so-called adiabatic basis, the Fourier modes $\widetilde{\psi}(\vec{q},t)$ read:
\begin{equation}
  \label{appa:quantization}
  \widetilde{\psi}(\vec{q},t)=\sum_{s}{\widetilde{U}_s(\vec{q},t)\widetilde{a}_s(\vec{q},t)+\widetilde{V}_s(-\vec{q},t)\widetilde{b}_s^\dagger(-\vec{q},t)}
  \ .
\end{equation}
The adiabatic four-spinors are chosen in close analogy to Eq.~(\ref{appa:spinoru1}) -- (\ref{appa:spinorv1}) such that they coincide with the vacuum solutions in the case of vanishing electric fields:
\begin{alignat}{3}
  \label{appa:spinoru2}
  \widetilde{U}_s(\vec{q},t)&\ = \ &\left(\gamma^0\widetilde{\omega}(\vec{q},t)-\vec{\gamma}\cdot\vec{\pi}(\vec{q},t)+m\right)\widetilde{G}^{(+)}(\vec{q},t)R_s& \ , \\
  \label{appa:spinorv2}
  \widetilde{V}_s(-\vec{q},t)&\ = \ &\left(-\gamma^0\widetilde{\omega}(\vec{q},t)-\vec{\gamma}\cdot\vec{\pi}(\vec{q},t)+m\right)\widetilde{G}^{(-)}(\vec{q},t)R_s& \ ,
\end{alignat}
with the adiabatic mode function $\widetilde{G}^{(\pm)}(\vec{q},t)$ given by:
\begin{equation}
  \label{appa:g_adiabatic}
  \widetilde{G}^{(\pm)}(\vec{q},t)=\frac{e^{\mp i\Theta(\vec{q},t_0,t)}}{\sqrt{2\widetilde{\omega}(\vec{q},t)(\widetilde{\omega}(\vec{q},t)\mp \pi_3(q_3,t))}} \ ,
\end{equation}
and the dynamical phase \colG{$\Theta(\vec{q},t_0,t)$} being defined as
\begin{equation}
  \label{appa:dynamical_phase}
  \Theta(\vec{q},t_0,t)=\int_{t_0}^{t}{\widetilde{\omega}(\vec{q},t')dt'} \ .
\end{equation}
The lower bound $t_0$ is not determined since it only fixes an arbitrary phase
at a given instant of time. Note that an interpretation in terms of
particles/antiparticles is only straightforward at asymptotic times when the
external electric field vanishes and the solutions
Eq.~(\ref{appa:g_adiabatic}) behave like plane waves.

In order to define the single-particle momentum distribution function $\widetilde{f}(\vec{q},t)$, we assume that we start with vacuum initial conditions at $t\to-\infty$:
\begin{equation}
  \bra{0}a_s^\dagger(\vec{q})a_s(\vec{q})\ket{0}=\bra{0}b_s^\dagger(\vec{q})b_s(\vec{q})\ket{0}=0 \ .
\end{equation}
We then define $\widetilde{f}(\vec{q},t)$ as the instantaneous quasi-particle number density for a given canonical momentum $\vec{q}$. Due to the absence of magnetic fields, we take the sum over both spin states, such that:
\begin{equation}
  \label{appa:dist_function} 
  \widetilde{f}(\vec{q},t)=\lim_{V\rightarrow\infty}\sum_{s=1,2}{\frac{\bra{0}\widetilde{a}^\dagger_s(\vec{q},t) \widetilde{a}_s(\vec{q},t)\ket{0}}{V}}=2|\widetilde{\beta}(\vec{q},t)|^2 \ .
\end{equation}
with $V$ being the (infinite) configuration space volume. As a consequence, the knowledge of $\widetilde{\beta}(\vec{q},t)$ allows for the calculation of $\widetilde{f}(\vec{q},t)$. In fact, the different representations of $\widetilde{\psi} (\vec{q},t)$ Eq.~(\ref{sec2b:quantization}) and Eq.~(\ref{appa:quantization}) translate into an expression for the Bogoliubov coefficients:
\begin{alignat}{4}
  \label{appa:bogoliubov_a}
  &\widetilde{\alpha}(\vec{q},t)& \ = \ & &i&\epsilon_\perp\widetilde{G}^{(-)}(\vec{q},t) \left[\partial_t-i\widetilde{\omega}(\vec{q},t)\right]\widetilde{g}^{(+)}(\vec{q},t)& \ , 
  \\
  \label{appa:bogoliubov_b}
  &\widetilde{\beta}(\vec{q},t)& \ = \ & -&i&\epsilon_\perp\widetilde{G}^{(+)}(\vec{q},t) \left[\partial_t+i\widetilde{\omega}(\vec{q},t)\right]\widetilde{g}^{(+)}(\vec{q},t)& \ ,
\end{alignat}
such that their time derivatives form an ODE system:
\begin{alignat}{6}
  &\frac{d}{dt}\widetilde{\alpha}(\vec{q},t)& \ = \ & &\frac{1}{2}\widetilde{Q}(\vec{q},t)&\widetilde{\beta}(\vec{q},t) e^{2i\Theta(\vec{q},t_0,t)}& \ ,
  \\
  &\frac{d}{dt}\widetilde{\beta}(\vec{q},t)& \ = \ &-&\frac{1}{2}\widetilde{Q}(\vec{q},t)&\widetilde{\alpha}(\vec{q},t) e^{-2i\Theta(\vec{q},t_0,t)}& \ ,
\end{alignat}
with $\widetilde{Q}(\vec{q},t)$ defined as in Eq.~(\ref{sec2b:q_function}). Introducing $\widetilde{\mathcal{C}}(\vec{q},t)=
2\widetilde{\alpha}(\vec{q},t)\widetilde{\beta}^*(\vec{q},t)$, this ODE system might be rewritten as:
\begin{alignat}{4}
  &\frac{d}{dt}\widetilde{\mathcal{C}}(\vec{q},t)& \ = \ & -\widetilde{Q}(\vec{q},t)\left[1-\widetilde{f}(\vec{q},t)\right]e^{-2i\Theta(\vec{q},t_0,t)}& \ , \\
  &\frac{d}{dt}\widetilde{f}(\vec{q},t)& \ = \ & -\widetilde{Q}(\vec{q},t)\operatorname{Re}\left[\widetilde{\mathcal{C}}(\vec{q},t)e^{2i\Theta(\vec{q},t_0,t)}\right]& \ ,
\end{alignat}
Formally integrating the first equation from a time of pure vacuum $t_\mathrm{vac}\to-\infty$ to $t$, yields the Vlasov equation for the single-particle momentum distribution function $\widetilde{f}(\vec{q},t)$ in its integro-differential form:
\colG{
\begin{eqnarray}
  \label{appa:qke_int}
  &\frac{d}{dt}\widetilde{f}(\vec{q},t)=  \widetilde{Q}(\vec{q},t)\int_{t_\mathrm{vac}}^{t}{dt'\widetilde{Q}(\vec{q},t')}\qquad \qquad \qquad \qquad\nonumber\\
  &\qquad \qquad \quad {\times}[1-\widetilde{f}(\vec{q},t')]\cos\left[2\Theta(\vec{q},t',t)\right],
\end{eqnarray}}
\noindent with $\widetilde{f}(\vec{q},t_\mathrm{vac})=0$. It is possible to rewrite this integro-differential equation in terms of an equivalent ODE system by introducing auxiliary functions $\widetilde{\rho}(\vec{q},t)$ and $\widetilde{\sigma}(\vec{q},t)$:
\begin{eqnarray}
  \label{appa:qkt_diff1}
  \frac{d}{dt}\widetilde{f}(\vec{q},t)&=&\widetilde{Q}(\vec{q},t)\,\widetilde{\rho}(\vec{q},t) \ , 
  \\
  \label{appa:qke_diff2}
  \frac{d}{dt}\widetilde{\rho}(\vec{q},t)&=&\widetilde{Q}(\vec{q},t)[1-\widetilde{f}(\vec{q},t)]-2\widetilde{\omega}(\vec{q},t)\,\widetilde{\sigma}(\vec{q},t) \ , \qquad
  \\
  \label{appa:qke_diff3}
  \frac{d}{dt}\widetilde{\sigma}(\vec{q},t)&=&2\widetilde{\omega}(\vec{q},t)\,\widetilde{\rho}(\vec{q},t) \ ,
\end{eqnarray}
with appropriate vacuum initial conditions $\widetilde{\rho}(\vec{q},t_\mathrm{vac})= \widetilde{\sigma}(\vec{q},t_\mathrm{vac})=0$.

\section{DHW functions for exactly solvable electric fields}
\label{appb}
In this appendix we give the analytic expressions for the DHW functions for the exactly solvable cases of the constant electric field and the Sauter-type electric field. The DHW functions are obtained as follows: We already derived the analytic expressions for the single-particle momentum distribution function $\widetilde{f}(\vec{q},t)$ in Sec.~\ref{sec2c}. As a consequence, according to Eq.~(\ref{sec2b:dhw_qkt1}) -- (\ref{sec2b:dhw_qkt3}) we are able to calculate the non-vanishing coefficients $\widetilde{\chi} ^{i=\{2,3\}}(\vec{q},t)$, \colG{cf. Eq.~(\ref{sec2b:basis_simplified})}, as well. Finally, we obtain the DHW functions according to Eq.~(\ref{sec2b:dhw_s}) -- (\ref{sec2b:dhw_t13}) after performing the phase space projection  Eq.~(\ref{sec2b:dhw_qkt}).

\subsection{Constant electric field}

In order to simplify the expression for the single-particle momentum distribution function $\widetilde{f}(u)$ derived in Eq.~(\ref{sec2c:1_dist}), we introduce the following abbreviations:
\begin{eqnarray}
  \widetilde{d}_1(u)&=&\left|D_{-1+i\eta/2}\left(-ue^{-i\frac{\pi}{4}}\right)\right|^2  
  \\
  \widetilde{d}_2(u)&=&\left|D_{i\eta/2}\left(-ue^{-i\frac{\pi}{4}}\right)\right|^2 
  \\
  \widetilde{d}_3(u)&=&e^{i\frac{\pi}{4}}D_{-1-i\eta/2}\left(-ue^{-i\frac{\pi}{4}}\right)\times\nonumber\\
  &&\qquad \qquad \quad D_{i\eta/2}\left(-ue^{-i\frac{\pi}{4}}\right) +c.c. \\
  \widetilde{d}_4(u)&=&e^{-i\frac{\pi}{4}}D_{-1-i\eta/2}\left(-ue^{-i\frac{\pi}{4}}\right)\times\nonumber\\
  &&\qquad \qquad \quad D_{i\eta/2}\left(-ue^{-i\frac{\pi}{4}}\right)  +c.c. 
\end{eqnarray}
which fulfill:
\begin{eqnarray}
  \partial_u\widetilde{d}_1(u)=&\quad \ \,  \widetilde{d}_4(u)&  \ , \\
  \partial_u\widetilde{d}_2(u)=&-\frac{\eta}{2}\widetilde{d}_4(u)& \ ,\\
  \partial_u\widetilde{d}_3(u)=&-u\,\widetilde{d}_4(u)&  \ .
\end{eqnarray}
We may then express $\widetilde{f}(u)$ in terms of $\widetilde{d}_{i}(u)$, such that $\widetilde{\chi}^{i=\{1,2,3\}}(u)$ are given by:
\begin{widetext}
\begin{eqnarray}
  \widetilde{\chi}^1(u)&=&1-e^{-\frac{\pi\eta}{4}}\left\{\frac{\eta}{2}\left(1-\frac{u}{\sqrt{2\eta+u^2}}\right)\widetilde{d}_1(u)+
  \left(1+\frac{u}{\sqrt{2\eta+u^2}}\right)\widetilde{d}_2(u)-\frac{\eta}{\sqrt{2\eta+u^2}}\widetilde{d}_3(u)\right\} , 
  \\
  \widetilde{\chi}^2(u)&=&\sqrt{\frac{\eta}{2}}\frac{1}{\sqrt{2\eta+u^2}}e^{-\frac{\pi\eta}{4}}
  \left\{-\eta\,\widetilde{d}_1(u)+2\widetilde{d}_2(u)+u\widetilde{d}_3(u)\right\} \ ,
  \\
  \widetilde{\chi}^3(u)&=&\frac{\sqrt{2\eta}}{(2\eta+u^2)^{3/2}}+\sqrt{\frac{\eta}{2}}\frac{1}{(2\eta+u^2)^{3/2}}
  e^{-\frac{\pi\eta}{4}}\left\{-\eta\,\widetilde{d}_1(u)-2\widetilde{d}_2(u)+(2\eta+u^2)^{3/2}\widetilde{d}_4(u)\right\} \ , 
\end{eqnarray}
\end{widetext}
with $\widetilde{\chi}^1(u)=1-\widetilde{f}(u)$. In order to obtain the DHW functions we perform the variable transformation $\vec{q}\to \vec{p}+e\vec{A}(t)$. Due to the linear relation between $q_3$ and $t$, this phase-space projection is trivial and reads:
\begin{eqnarray}
  &u\to\sqrt{\frac{2}{\epsilon}}\hat{p}_3 \quad \mathrm{and} \quad d_i(\vec{p})=\widetilde{d}_i\left(\sqrt{\frac{2}{\epsilon}}\hat{p}_3\right)\ , & 
\end{eqnarray}
where we introduced the dimensionless phase-space kinetic momentum $\hat{p}_3=\frac{p_3}{m}$. Note that the $d_i(\vec{p})$ implicitly depend on the orthogonal kinetic momentum $\vec{p}_\perp$ by means of $\eta=\epsilon_\perp^2/eE_0$ with $\epsilon_\perp^2=m^2+\vec{p}_\perp^2$. Obviously, $d_i(\vec{p})$ do not depend on the time variable $t$ but only on the kinetic momentum $\vec{p}$ such that the Schwinger effect in a constant electric field might be regarded as a quasi-static problem. The phase-space coefficients $\chi^{i=\{1,2,3\}}(\vec{p})$ which allow for the calculation of the DHW functions Eq.~(\ref{sec2b:dhw_s}) -- (\ref{sec2b:dhw_t13}) \colG{thus} read:
\begin{widetext}
\begin{eqnarray}
  \chi^1(\vec{p})&=&1-e^{-\frac{\pi\epsilon_\perp^2}{4eE_0}}\left\{\frac{\epsilon_\perp^2}{2eE_0}\left(1-\frac{p_3}{\omega(\vec{p})}\right)d_1(\vec{p})+
  \left(1+\frac{p_3}{\omega(\vec{p})}\right)d_2(\vec{p})-\frac{\epsilon_\perp^2}{\sqrt{2eE_0}\omega(\vec{p})}d_3(\vec{p})\right\} \ , 
  \\
  \label{appb:1_solution_coeff}
  \chi^2(\vec{p})&=&\frac{\epsilon_\perp}{2\omega(\vec{p})}e^{-\frac{\pi\epsilon_\perp^2}{4eE_0}}
  \left\{-\frac{\epsilon_\perp^2}{eE_0}\,d_1(\vec{p})+2d_2(\vec{p})+\sqrt{\frac{2}{eE_0}}p_3d_3(\vec{p})\right\} \ ,
  \\
  \chi^3(\vec{p})&=&\frac{eE_0\epsilon_\perp}{2\omega^3(\vec{p})}\left(1+\frac{1}{2}e^{-\frac{\pi\epsilon_\perp^2}{4eE_0}}
  \left\{-\frac{\epsilon_\perp^2}{eE_0}\,d_1(\vec{p})-2d_2(\vec{p})+\left(\frac{2}{eE_0}\right)^{3/2}\omega^3(\vec{p})d_4(\vec{p})\right\}\right) \ . 
\end{eqnarray}
\end{widetext}

\subsection{Sauter-type electric field}
We start from the expression for the single-particle momentum distribution function $\widetilde{f}(\hat{q}_3,u)$ given in Eq.~(\ref{sec2c:2_dist}) and introduce the following abbreviations:

\begin{eqnarray}
  \widetilde{h}_1(\hat{q}_3,u)&=&\left|F(\tilde{a},\tilde{b},\tilde{c};u)\right|^2 \\
  \widetilde{h}_2(\hat{q}_3,u)&=&\left|\tfrac{\tilde{a}\tilde{b}}{\tilde{c}}F(1+\tilde{a},1+\tilde{b},1+\tilde{c};u)\right|^2 \\
  \widetilde{h}_3(\hat{q}_3,u)&=&-i\tfrac{\tilde{a}\tilde{b}}{\tilde{c}}F(1+\tilde{a},1+\tilde{b},1+\tilde{c};u)\times\qquad \quad \nonumber\\
  && \qquad \qquad \qquad  F(\tilde{a}^*,\tilde{b}^*,\tilde{c}^*;u)+c.c. \quad
\end{eqnarray}
and
\begin{eqnarray}
  \widetilde{\varrho}_1(\hat{q}_3,u)&=& [\widehat{\omega}(\hat{q}_3,u)-(1-u)\widehat{\omega}(\hat{q}_3,0)-u\widehat{\omega}(\hat{q}_3,1)]^2 \quad \quad \ \\
  \widetilde{\varrho}_2(\hat{q}_3,u)&=& 4\gamma^2\epsilon^2u^2(1-u)^2 \\
   \widetilde{\varrho}_3(\hat{q}_3,u)&=& 2\gamma\epsilon u(1-u) \times\nonumber\\
  &&[\widehat{\omega}(\hat{q}_3,u)-(1-u)\widehat{\omega}(\hat{q}_3,0)-u\widehat{\omega}(\hat{q}_3,1)]
\end{eqnarray}
such that $\widetilde{f}(\hat{q}_3,u)$ can be written as:
\begin{equation}
  \label{appb:2_dist}
  \widetilde{f}(\hat{q}_3,u)=\widetilde{N}_f(\hat{q}_3)\left(1+\frac{\widehat{\pi}_3(\hat{q}_3,u)}{\widehat{\omega}(\hat{q}_3,u)}\right) \sum_{i=1}^{3}{\widetilde{\varrho}_i(\hat{q}_3,u)\widetilde{h}_i(\hat{q}_3,u)} .
\end{equation}
\begin{widetext}
Again note that $\{\tilde{a},\tilde{b}, \tilde{c}\}$, which have been defined in Eq.~(\ref{sec2c:2_arguments}), only depend on $\hat{q}_3$ but not on $u$. Taking into account the general derivation formula for the Gauss hypergeometric \colG{function} Eq.~(\ref{sec2c:2_hypergeometric}), we can explicitly calculate the first and second derivative of Eq.~(\ref{appb:2_dist}). After calculating $\partial_u\widetilde{f}(\hat{q}_3,u)$ and $\partial_u^2\widetilde{f}(\hat{q}_3,u)$, we are able to determine the coefficients $\widetilde{\chi}^{i=\{1,2,3\}}(\hat{q}_3,u)$ according to:
\begin{eqnarray}
  &\widetilde{\chi}^1(\hat{q}_3,u)=1-\widetilde{f}(\hat{q}_3,u) \qquad \qquad \qquad \qquad \qquad \qquad \qquad \qquad \qquad \qquad \qquad \qquad \qquad \qquad \quad \ \\
  &\widetilde{\chi}^2(\hat{q}_3,u)=\frac{\gamma\widehat{\omega}^2(\hat{q}_3,u)}{2\sqrt{1+\hat{\kappa}^2}}\partial_u\widetilde{f}(\hat{q}_3,u) \qquad \qquad \qquad \qquad \qquad \qquad \qquad \qquad \qquad \qquad \qquad \qquad \qquad \ \\
  &\widetilde{\chi}^3(\hat{q}_3,u)=\frac{2\epsilon\sqrt{1+\hat{\kappa}^2}
  u(1-u)}{\widehat{\omega}^3(\hat{q}_3,u)}\left\{1-\widetilde{f}(\hat{q}_3,u)-\frac{\gamma\widehat{\omega}^2(\hat{q}_3,u)\widehat{\pi}_3(\hat{q}_3,u)}{1+\hat{\kappa}^2}\partial_u\widetilde{f}(\hat{q}_3,u)-\frac{\gamma^2\widehat{\omega}^4(\hat{q}_3,u)}{4(1+\hat{\kappa}^2)}\partial_u^2\widetilde{f}(\hat{q}_3,u)\right\} 
\end{eqnarray}
\end{widetext}
\newpage
In order to obtain the coefficients in phase-space, we have to perform the variable transformation $\vec{q}\to \vec{p}+e\vec{A}(t)$, which reads:
\begin{eqnarray}
  \hat{q}_3\to\hat{p}_3-\tfrac{2u-1}{\gamma}
\end{eqnarray}
As a consequence, the quantities $\widehat{\omega}(\hat{q}_3,u)$ and $\widehat{\pi}_3(\hat{q}_3,u)$ only depend on $\hat{p}_3$ but not on $u$ \colG{after performing} this variable transformation:
\begin{eqnarray}
   \widehat{\pi}_3(\hat{q}_3,u)&\to&\hat{p}_3 \ , \\ 
   \widehat{\omega}(\hat{q}_3,u)&\to&\hat{\omega}(\hat{p}_3)=\sqrt{1+\hat{\kappa}^2+\hat{p}_3^2} \ ,
\end{eqnarray}
On the other hand, any function of the canonical momentum $\hat{q}_3$ only,
e.g. $\widehat{\omega}(\hat{q}_3,0)$ or $\widehat{\omega}(\hat{q}_3,1)$,
\colG{acquires {a} dependence} on both the phase-space kinetic momentum $\hat{p}_3$ and the time variable $u$:
\begin{eqnarray}
  \widehat{\pi}_{3}(\hat{q}_3,0)=&\hat{q}_3-\frac{1}{\gamma}\to
  \hat{p}_3-\frac{2u}{\gamma} \ , \quad& \\ 
  \widehat{\pi}_{3}(\hat{q}_3,1)=&\hat{q}_3+\frac{1}{\gamma}\to
  \hat{p}_3-\frac{2u-2}{\gamma} \ , &
\end{eqnarray}
and

\begin{eqnarray}
  \widehat{\omega}(\hat{q}_3,0)\to&
  \sqrt{1+\hat{\kappa}^2+\left(\hat{p}_3-\frac{2u}{\gamma}\right)^2} \ , \quad& \\
  \widehat{\omega}(\hat{q}_3,1)\to&
  \sqrt{1+\hat{\kappa}^2+\left(\hat{p}_3-\frac{2u-2}{\gamma}\right)^2} \  &
\end{eqnarray}
Therefore, \colG{whereas} the functions $\widetilde{h}_i(\hat{q}_3,u)$
\colG{depend on $u$ solely} through the last argument of the Gauss
hypergeometric \colG{function}, the transformed functions $h_i(\hat{p}_3;u)$
have a twofold $u$ dependence: On the one hand, there is still the $u$
dependence due to the last argument. On the other hand, due to the fact that
the parameters Eq.~(\ref{sec2c:2_arguments}) were function of $\hat{q}_3$
only, they will depend on both $\hat{p}_3$ and $u$ after the transformation to
phase space.

\end{appendix}

\end{document}